\documentclass[12pt]{article}
\usepackage{axodraw,epsfig,cite}
\input paperdef

\evensidemargin \oddsidemargin
\def\_{\rule{.3em}{.15ex}}

\def\slash#1{\setbox0=\hbox{$#1$}#1\hskip-\wd0\dimen0=5pt\advance
       \dimen0 by-\ht0\advance\dimen0 by\dp0\lower0.5\dimen0\hbox
         to\wd0{\hss\sl/\/\hss}}

\def\ra{\rightarrow}

\begin{document}
\thispagestyle{empty}

\def\thefootnote{\fnsymbol{footnote}}

\begin{minipage}[t]{5cm}
BNL--HET--00/37\\
CERN--TH/2000--369\\
DESY 00--139
\end{minipage}
\hfill
\begin{minipage}[t]{4cm}
KA--TP--6--2000\\
TUM--HEP--392/00\\
LC--TH--2001--037\\
hep-ph/0102081
\end{minipage}

%\begin{flushright}
%October 2000
%\end{flushright}

\vspace{0.5cm}

\begin{center}

{\large\sc {\bf Neutral MSSM Higgs-boson production at \epem\ colliders}}

\vspace{0.2cm}

{\large\sc {\bf in the Feynman-diagrammatic approach}}
 
\vspace{0.4cm}

{\sc 
S.~Heinemeyer$^{1}$%
\footnote{email: Sven.Heinemeyer@bnl.gov}%
, W.~Hollik$^{2}$%
\footnote{email: Wolfgang.Hollik@physik.uni-karlsruhe.de}%
, J.~Rosiek$^{3}$%
\footnote{email: Janusz.Rosiek@ph.tum.de}%
~and G.~Weiglein$^{4}$%
\footnote{email: Georg.Weiglein@cern.ch}
}

\vspace*{0.5cm}

{\sl
%$^1$DESY Theorie, Notkestr. 85, 22603 Hamburg, Germany
$^1$HET, Physics Department, Brookhaven Natl.\ Lab., Upton, NY
11973, USA

\vspace*{0.2cm}

$^2$Institut f\"ur Theoretische Physik, Universit\"at Karlsruhe, \\
D-76128 Karlsruhe, Germany

\vspace*{0.2cm}

$^3$Physik-Department, Technische Universit\"at M\"unchen, \\
D-85747 Garching, Germany\\
and\\
Institute of Theoretical Physics, Warsaw University, \\
PL-00681 Warsaw, Poland\\
%$^3$Physik-Department, Technische Universit\"at M\"unchen, \\
%D-85747 Garching, Germany

\vspace*{0.2cm}

$^4$CERN, TH Division, CH-1211 Geneva 23, Switzerland
}

\end{center}

\vspace*{0.2cm}

\begin{abstract}
~We calculate the cross sections for the neutral Higgs-boson
production at \epem~colliders in the Minimal Supersymmetric Standard
Model (MSSM) using the Feynman-diagrammatic approach and the
on-shell renormalization scheme.  We incorporate the Higgs-boson
propagator corrections, evaluated up to \twol\ order, into the
prediction of the cross sections for the Higgs-boson production
mechanism \eetohZhA.  The propagator corrections consist of the full
\onel\ contribution, including the effects of non-vanishing external
momentum, and at the \twol\ level of the dominant corrections of
$\oaas$ and further sub-dominant contributions. The results are
supplemented with the complete set of \onel\ vertex and box
corrections.  The effects of the \twol\ propagator corrections are
investigated in detail.  We briefly discuss also the effect of the box
contributions for high $\sqrt{s}$.  We compare our results with the
case where only the corrections to the effective mixing angle, 
evaluated within the renormalization-group-improved one-loop Effective
Potential approach, are taken into account. 
We find agreement better than 10\% for 
LEP2 energies and deviations larger than 20\% for $\sqrt{s} = 500$
GeV.  
\end{abstract}
%\pacs{}

\def\thefootnote{\arabic{footnote}}
\setcounter{page}{0}
\setcounter{footnote}{0}

\newpage

%%%%%%%%%%%%%%%%%%%%%%%%%%%%%%%%%%%%%%%%%%%%%%%%%%%%%%%%%%%%%%
%%%%%%%%%%%%%%%%%%%%%%%%%%%%%%%%%%%%%%%%%%%%%%%%%%%%%%%%%%%%%%

\section{Introduction}

The search for the light neutral Higgs boson is a crucial test of
Supersymmetry that can be performed with the present and the next
generation of accelerators.  The prediction of a relatively light
Higgs boson is common to all supersymmetric models whose couplings
remain in the perturbative regime up to a very high energy
scale~\cite{susylighthiggs}.  Finding the Higgs boson is thus one of
the main goals of high-energy physics. 
Possible indications for a Higgs boson with a mass around $115 \gev$
have recently been observed in the LEP analysis~\cite{lephiggs}.
As a second step, once a
scalar particle has been found, it is necessary to measure the
dominant production cross sections as well as the decay widths and
branching ratios of the main decay channels to a high accuracy. Such
precise measurements of Higgs--gauge-boson and Higgs--Yukawa couplings
allow to experimentally investigate the details of the Higgs
mechanism.  Finally, also the Higgs self-couplings will have to be
measured in order to reconstruct the Higgs potential. 
A future linear \epem\ collider with high luminosity~\cite{teslacdr}
can provide a sufficiently clean environment to measure both 
the Higgs couplings to other particles as
well as its self couplings~\cite{hhh} with high
precision, allowing to distinguish
between a standard and a non-standard Higgs boson.  In this paper we
concentrate on the production of the neutral Higgs bosons of the
Minimal Supersymmetric Standard Model (MSSM) in \epem\ annihilations
and provide results for the production cross sections for general
model parameters, together with a detailed numerical study.

In the MSSM, the mass of the lightest Higgs boson, $\Mh$, is bounded
from above by $\Mh \lsim 135 \gev$, including radiative corrections up
to \twol\
order~\cite{mhiggs1l,mhiggsRG1a,mhiggsRG1b,mhiggsRG2,mhiggsEP1,mhiggsEP2,mhiggsletter,mhiggslong,maulpaul}.
The most promising channels for the production of the supersymmetric
neutral Higgs particles 
at LEP2 energies and in the first phase of future $e^+e^-$ colliders 
are the Higgs-strahlung process~\cite{hprod}
\BE 
e^+e^- \to Z\,h\,(H) , 
\EE
and the associated production of a scalar and a pseudoscalar Higgs boson 
\BE
e^+e^- \to A\,h\,(H)~.  
\EE 

We compute the MSSM predictions for the cross sections of both
channels in the Feynman-diagrammatic (FD) approach using the on-shell
renormalization scheme. 
We take into account the complete set of \onel\ contributions,
thereby keeping the full dependence on all kinematical variables.
The one-loop contributions consist of the corrections to the Higgs- 
and gauge-boson propagators, where the former contain the dominant 
electroweak \onel\ corrections of
\order{\gf\mt^4}, and of the contributions to the 3-point and 4-point vertex
functions~\cite{CPRlet,CPR,mhiggsf1l,boxes,higgsprod}.
We combine the complete one-loop result with the dominant
two-loop QCD corrections of 
\order{\gf\als\mt^4}~\cite{mhiggsletter,mhiggslong} and further
sub-dominant corrections.  In this way the currently most
accurate results for the cross sections are obtained.  

Furthermore
we show analytically that the Higgs-boson propagator corrections with
neglected momentum dependence can be absorbed into the tree-level
coupling using the effective mixing angle from the neutral $\cp$-even
Higgs-boson sector. We  compare our results for the
cross sections with the approximation in which only the corrections to 
the effective mixing angle, evaluated within the
renormalization-group-improved one-loop 
Effective Potential approach, are taken 
into account.
For most parts of the MSSM parameter space we find
agreement better than 10\% for the highest LEP energies,
while for $\sqrt{s} = 500 \gev$ the difference can reach 25\%.

The paper is organized as follows: in section~\ref{sec:form} the basic
formulae are presented and it is shown analytically how the
Higgs-boson-propagator corrections with neglected external momenta are
related to the effective mixing angle in the Effective Potential
approach (EPA).  Section~\ref{sec:num} contains the numerical analysis
for the production cross sections and the comparison of our full
result with the EPA.
The conclusions can be found in
section~\ref{sec:concl}.

%%%%%%%%%%%%%%%%%%%%%%%%%%%%%%%%%%%%%%%%%%%%%%%%%%%%%%%%%%%%%%
%%%%%%%%%%%%%%%%%%%%%%%%%%%%%%%%%%%%%%%%%%%%%%%%%%%%%%%%%%%%%%

\section{Cross sections for Higgs-particle production in 
         $e^+e^-$ collisions}
\label{sec:form}

\subsection{Classification of radiative corrections}
\label{subsec:rcXsec}

The two Higgs-field doublets giving rise to electroweak symmetry
breaking within the MSSM accommodate five physical Higgs bosons~\cite{hhg}.
At the tree-level, two input parameters (besides the
parameters of the Standard Model (SM) gauge-sector) 
are needed to describe the Higgs
sector.  We choose them to be $\tb$, the ratio of the two vacuum
expectation values, and $\MA$, the mass of the $\cp$-odd Higgs boson.
The $\cp$-even neutral mass eigenstates are obtained from the
interaction eigenstates by the rotation 
\BE
%\Hi = 
\VL H_1 \\ H_2 \VR \equiv
\VL H \\ h \VR = \ML \Ca & \Sa \\ -\Sa & \Ca \MR 
\VL \phi_1^0 \\ \phi_2^0 \VR 
\label{higgsrotation}
\EE
with the tree-level mixing angle $\alpha$ related to
$\tb$, $\MA$ and $\MZ$ by
\BE
\tan 2\alpha = \tan 2\beta \frac{\MA^2 + \MZ^2}{\MA^2 - \MZ^2},
\quad - \frac{\pi}{2} < \alpha < 0.
%\frac{\pi}{2}. 
\EE

%%%%%%%%%%%%%%%%%%%%%%%%%%%%%%%%%%%%%%%%%%%%%%%%%%%%%%%%%%%%%%
%%%%%%%%%%%%%%%%%%%%%%%%%%%%%%%%%%%%%%%%%%%%%%%%%%%%%%%%%%%%%%

\bigskip
Two main sources for the production of supersymmetric Higgs particles
in $e^+e^-$ collisions are the
Higgs-strahlung process~\cite{hprod}
\BE
e^+e^- \to Z\Hi,~i=1,2
\label{eetohZ}
\EE 
(using the compact notation of \refeq{higgsrotation}) and the
associated production of scalar and pseudoscalar Higgs bosons,
\BE 
e^+e^- \to A\Hi,~i=1,2 .
\label{eetohA}
\EE
We do not discuss here the possibility of Higgs-particle
production by brems\-strahlung off heavy quarks (e.g. $e^+e^- \to
\bar{b}b\Hi$, which can be significant for large $\tb$~\cite{hbb}), or
by $W^+W^-$ fusion, which becomes important for center of mass
system (CMS) energies of ${\cal O}(500$~GeV)~\cite{hprod}.

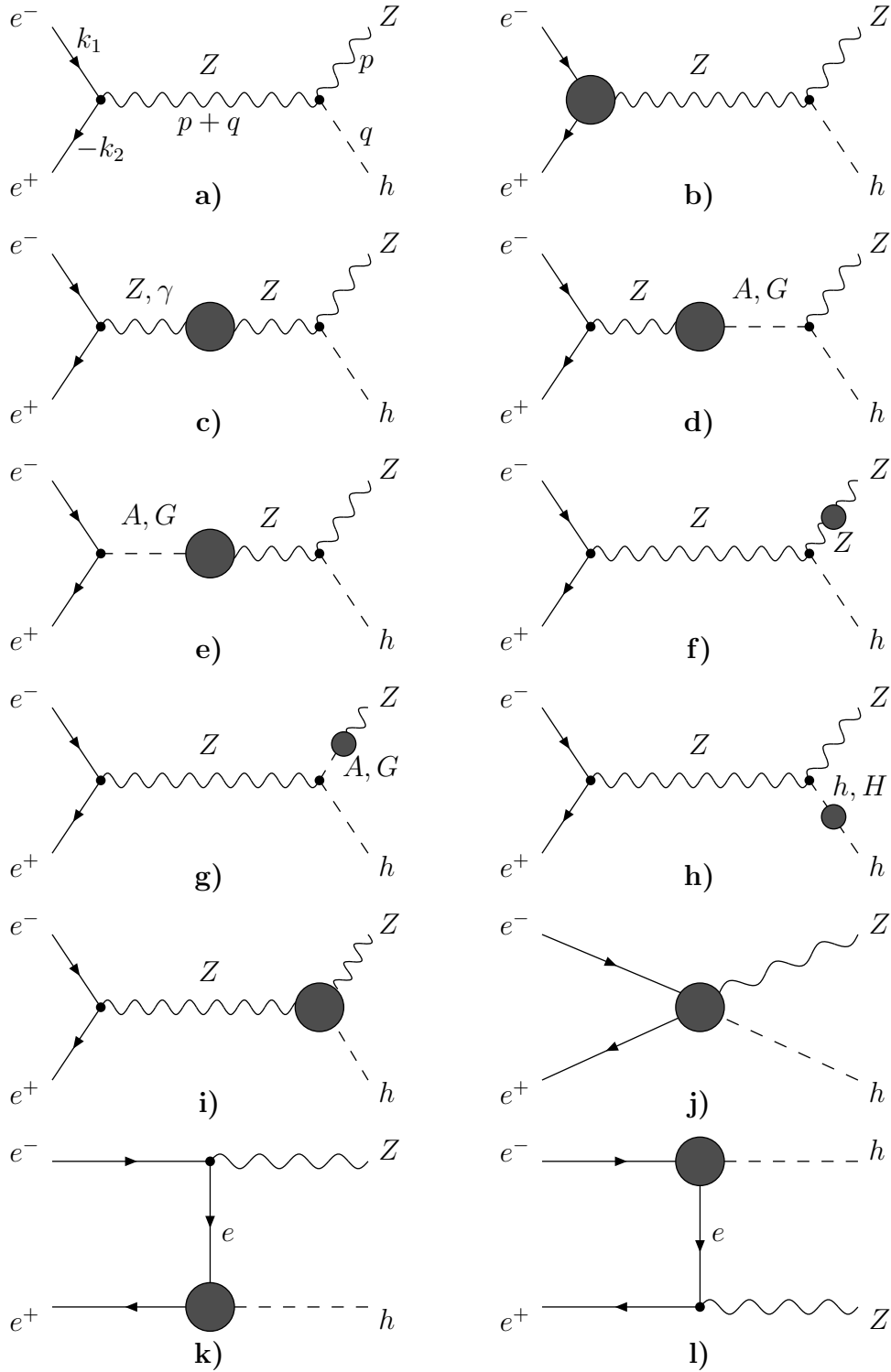
\begin{figure}[ht!]
\begin{center}
\begin{tabular}{lp{1cm}r}
\begin{picture}(150,75)(0,0)
% left lower skew line
\ArrowLine(30,40)(10,10)
\Text(5,5)[r]{\mbox{$e^+$}}
\Text(20,20)[l]{\mbox{$-k_2$}}
% left upper skew line
\ArrowLine(10,70)(30,40)
\Text(5,75)[r]{\mbox{$e^-$}}
\Text(20,65)[l]{\mbox{$k_1$}}
% left vertex
\Vertex(30,40){2}
% Z0 line
\Photon(30,40)(120,40){3}{8}
\Text(75,55)[c]{\mbox{$Z$}}
\Text(75,30)[c]{\mbox{$p+q$}}
% right vertex
\Vertex(120,40){2}
% right lower skew line
\DashLine(120,40)(140,10){5}
\Text(145,5)[l]{\mbox{$h$}}
\Text(137,25)[l]{\mbox{$q$}}
% right upper skew line
\Photon(120,40)(140,70){3}{3}
\Text(145,75)[l]{\mbox{$Z$}}
\Text(137,55)[l]{\mbox{$p$}}
% diagram symbol
\Text(75,0)[c]{\mbox{\bf a)}}
\end{picture}
&&
\begin{picture}(150,75)(0,0)
% left lower skew line
\ArrowLine(30,40)(10,10)
\Text(5,5)[r]{\mbox{$e^+$}}
% left upper skew line
\ArrowLine(10,70)(30,40)
\Text(5,75)[r]{\mbox{$e^-$}}
% left vertex
\GCirc(30,40){10}{0.3}                                                      
% Z0 line
\Photon(40,40)(120,40){3}{8}
\Text(75,55)[c]{\mbox{$Z$}}
% right vertex
\Vertex(120,40){2}
% right lower skew line
\DashLine(120,40)(140,10){5}
\Text(145,5)[l]{\mbox{$h$}}
% right upper skew line
\Photon(120,40)(140,70){3}{3}
\Text(145,75)[l]{\mbox{$Z$}}
% diagram symbol
\Text(75,0)[c]{\mbox{\bf b)}}
\end{picture}
\\
\\
\begin{picture}(150,75)(0,0)
% left lower skew line
\ArrowLine(30,40)(10,10)
\Text(5,5)[r]{\mbox{$e^+$}}
% left upper skew line
\ArrowLine(10,70)(30,40)
\Text(5,75)[r]{\mbox{$e^-$}}
% left vertex
\Vertex(30,40){2}
% Z0 line
\Photon(30,40)(120,40){3}{8}
\GCirc(75,40){10}{0.3}                                                      
\Text(50,55)[c]{\mbox{$Z,\gamma$}}
\Text(100,55)[c]{\mbox{$Z$}}
% right vertex
\Vertex(120,40){2}
% right lower skew line
\DashLine(120,40)(140,10){5}
\Text(145,5)[l]{\mbox{$h$}}
% right upper skew line
\Photon(120,40)(140,70){3}{3}
\Text(145,75)[l]{\mbox{$Z$}}
% diagram symbol
\Text(75,0)[c]{\mbox{\bf c)}}
\end{picture}
&&
\begin{picture}(150,75)(0,0)
% left lower skew line
\ArrowLine(30,40)(10,10)
\Text(5,5)[r]{\mbox{$e^+$}}
% left upper skew line
\ArrowLine(10,70)(30,40)
\Text(5,75)[r]{\mbox{$e^-$}}
% left vertex
\Vertex(30,40){2}
% Z0 line
\Photon(30,40)(75,40){3}{4}
\Text(50,55)[c]{\mbox{$Z$}}
\GCirc(75,40){10}{0.3}                                                      
\DashLine(85,40)(120,40){5}
\Text(100,55)[c]{\mbox{$A,G$}}
% right vertex
\Vertex(120,40){2}
% right lower skew line
\DashLine(120,40)(140,10){5}
\Text(145,5)[l]{\mbox{$h$}}
% right upper skew line
\Photon(120,40)(140,70){3}{3}
\Text(145,75)[l]{\mbox{$Z$}}
% diagram symbol
\Text(75,0)[c]{\mbox{\bf d)}}
\end{picture}
\\
\\
\begin{picture}(150,75)(0,0)
% left lower skew line
\ArrowLine(30,40)(10,10)
\Text(5,5)[r]{\mbox{$e^+$}}
% left upper skew line
\ArrowLine(10,70)(30,40)
\Text(5,75)[r]{\mbox{$e^-$}}
% left vertex
\Vertex(30,40){2}
% Z0 line
\Photon(75,40)(120,40){3}{4}
\Text(100,55)[c]{\mbox{$Z$}}
\GCirc(75,40){10}{0.3}                                                      
\DashLine(30,40)(65,40){5}
\Text(50,55)[c]{\mbox{$A,G$}}
% right vertex
\Vertex(120,40){2}
% right lower skew line
\DashLine(120,40)(140,10){5}
\Text(145,5)[l]{\mbox{$h$}}
% right upper skew line
\Photon(120,40)(140,70){3}{3}
\Text(145,75)[l]{\mbox{$Z$}}
% diagram symbol
\Text(75,0)[c]{\mbox{\bf e)}}
\end{picture}
&&
\begin{picture}(150,75)(0,0)
% left lower skew line
\ArrowLine(30,40)(10,10)
\Text(5,5)[r]{\mbox{$e^+$}}
% left upper skew line
\ArrowLine(10,70)(30,40)
\Text(5,75)[r]{\mbox{$e^-$}}
% left vertex
\Vertex(30,40){2}
% Z0 line
\Photon(30,40)(120,40){3}{8}
\Text(75,55)[c]{\mbox{$Z$}}
% right vertex
\Vertex(120,40){2}
% right lower skew line
\DashLine(120,40)(140,10){5}
\Text(145,5)[l]{\mbox{$h$}}
% right upper skew line
\Photon(120,40)(127.5,52.5){3}{1.5}
\Photon(132.5,57.5)(140,70){3}{1.5}
\GCirc(130,55){5}{0.3}                                                      
\Text(145,75)[l]{\mbox{$Z$}}
\Text(130,45)[l]{\mbox{$Z$}}
% diagram symbol
\Text(75,0)[c]{\mbox{\bf f)}}
\end{picture}
\\
\\
\begin{picture}(150,75)(0,0)
% left lower skew line
\ArrowLine(30,40)(10,10)
\Text(5,5)[r]{\mbox{$e^+$}}
% left upper skew line
\ArrowLine(10,70)(30,40)
\Text(5,75)[r]{\mbox{$e^-$}}
% left vertex
\Vertex(30,40){2}
% Z0 line
\Photon(30,40)(120,40){3}{8}
\Text(75,55)[c]{\mbox{$Z$}}
% right vertex
\Vertex(120,40){2}
% right lower skew line
\DashLine(120,40)(140,10){5}
\Text(145,5)[l]{\mbox{$h$}}
% right upper skew line
\Photon(132.5,57.5)(140,70){3}{1.5}
%\Photon(135,55)(140,70){3}{1}
\DashLine(120,40)(127.5,52.5){5}
\GCirc(130,55){5}{0.3}                                                      
\Text(145,75)[l]{\mbox{$Z$}}
\Text(130,45)[l]{\mbox{$A,G$}}
% diagram symbol
\Text(75,0)[c]{\mbox{\bf g)}}
\end{picture}
&&
\begin{picture}(150,75)(0,0)
% left lower skew line
\ArrowLine(30,40)(10,10)
\Text(5,5)[r]{\mbox{$e^+$}}
% left upper skew line
\ArrowLine(10,70)(30,40)
\Text(5,75)[r]{\mbox{$e^-$}}
% left vertex
\Vertex(30,40){2}
% Z0 line
\Photon(30,40)(120,40){3}{8}
\Text(75,55)[c]{\mbox{$Z$}}
% right vertex
\Vertex(120,40){2}
% right lower skew line
\DashLine(120,40)(128.5,27.5){5}
\DashLine(131.5,22.5)(140,10){5}
\GCirc(130,25){5}{0.3}                                                      
\Text(145,5)[l]{\mbox{$h$}}
\Text(130,37)[l]{\mbox{$h,H$}}
% right upper skew line
\Photon(120,40)(140,70){3}{3}
\Text(145,75)[l]{\mbox{$Z$}}
% diagram symbol
\Text(75,0)[c]{\mbox{\bf h)}}
\end{picture}
\\
\\
\begin{picture}(150,75)(0,0)
% left lower skew line
\ArrowLine(30,40)(10,10)
\Text(5,5)[r]{\mbox{$e^+$}}
% left upper skew line
\ArrowLine(10,70)(30,40)
\Text(5,75)[r]{\mbox{$e^-$}}
% left vertex
\Vertex(30,40){2}
% Z0 line
\Photon(30,40)(120,40){3}{8}
\Text(75,55)[c]{\mbox{$Z$}}
% right vertex
\GCirc(120,40){10}{0.3}                                                      
% right lower skew line
\DashLine(128,30)(140,10){5}
\Text(145,5)[l]{\mbox{$h$}}
% right upper skew line
\Photon(127.5,47.5)(140,70){3}{3}
\Text(145,75)[l]{\mbox{$Z$}}
% diagram symbol
\Text(75,0)[c]{\mbox{\bf i)}}
\end{picture}
&&
\begin{picture}(150,75)(0,0)
% left lower skew line
\ArrowLine(67,36)(10,10)
\Text(5,5)[r]{\mbox{$e^+$}}
% left upper skew line
\ArrowLine(10,70)(67,46)
\Text(5,75)[r]{\mbox{$e^-$}}
% Box & vertices
\GCirc(75,40){10}{0.3}                                                      
% right lower skew line
\DashLine(86,35)(140,10){5}
\Text(145,5)[l]{\mbox{$h$}}
% right upper skew line
\Photon(83,46)(140,70){3}{3}
\Text(145,75)[l]{\mbox{$Z$}}
% diagram symbol
\Text(75,0)[c]{\mbox{\bf j)}}
\end{picture}
\\
\\
\begin{picture}(150,75)(0,0)
% left lower line
\ArrowLine(75,10)(10,10)
\Text(5,5)[r]{\mbox{$e^+$}}
% left upper skew line
\ArrowLine(10,70)(75,70)
\Text(5,75)[r]{\mbox{$e^-$}}
% upper vertex
\GCirc(75,10){10}{0.3}                                                      
% Z0 line
\ArrowLine(75,70)(75,20)
\Text(80,40)[l]{\mbox{$e$}}
% lower vertex
\Vertex(75,70){2}
% right lower skew line
\DashLine(85,10)(140,10){5}
\Text(145,5)[l]{\mbox{$h$}}
% right upper skew line
\Photon(75,70)(140,70){3}{4}
\Text(145,75)[l]{\mbox{$Z$}}
% diagram symbol
\Text(75,-10)[c]{\mbox{\bf k)}}
\end{picture}
&&
\begin{picture}(150,75)(0,0)
% left lower line
\ArrowLine(75,10)(10,10)
\Text(5,5)[r]{\mbox{$e^+$}}
% left upper skew line
\ArrowLine(10,70)(75,70)
\Text(5,75)[r]{\mbox{$e^-$}}
% upper vertex
\GCirc(75,70){10}{0.3}                                                      
% Z0 line
\ArrowLine(75,60)(75,10)
\Text(80,40)[l]{\mbox{$e$}}
% lower vertex
\Vertex(75,10){2}
% right lower skew line
\Photon(75,10)(140,10){3}{4}
\Text(145,5)[l]{\mbox{$Z$}}
% right upper skew line
\DashLine(85,70)(140,70){5}
\Text(145,75)[l]{\mbox{$h$}}
% diagram symbol
\Text(75,-10)[c]{\mbox{\bf l)}}
\end{picture}
\\
\\
\end{tabular}
\caption{Generic diagrams contributing to the $e^+e^-\ra Zh$ cross section.}
\label{fig:diagrams}
\end{center}
\end{figure}

The set of diagrams taken into account for Higgs-strahlung \eetohZ\ is
schematically shown in \reffi{fig:diagrams}, where a) is the tree-level
diagram. The shaded blobs
summarize the loops with all possible virtual particles, except
photons in the $Ze^+e^-$ vertex corrections%
\footnote{
These virtual
IR-divergent photonic corrections constitute, together with
real-photon bremsstrahlung, the initial-state QED corrections, which
are conventionally treated separately and are the same as for the
SM Higgs-boson production.
}%
.  More details can be found in
\citeres{CPR,boxes,higgsprod}.  An analogous set has been evaluated
for the second process, \eetohA.

%%%%%%%%%%%% F I G U R E %%%%%%%%%%%%%%%%%%%%%%%%%%%%%%%%%%%%%

%%%%%%%%%%%% F I G U R E %%%%%%%%%%%%%%%%%%%%%%%%%%%%%%%%%%%%%

For completeness, in \reffi{fig:diagrams} also contributions are
shown (e.g.\ the $A,G$--$Z$ mixing contributions and the longitudinal
parts of the $Z$ and $\ga$--$Z$ self-energies) that are proportional to 
the electron mass or vanish completely after contraction with the
polarization vector of the $Z$~boson.
The different types of corrections can be summarized as follows:
\begin{itemize}
\item[(i)] Corrections to the $e$, $Z$, $\ga$ and $\ga$--$Z$
  self-energies on the internal and external lines and to the (initial
  state) $Ze^+e^-$ and $\ga e^+e^-$ vertices, b) -- g).
\item[(ii)] Corrections to the scalar and pseudoscalar propagators, h).
\item[(iii)] Corrections to the $ZZH_i$ ($ZAH_i$) vertex, i).
\item[(iv)] Box-diagram contributions and $t$-channel-exchange diagrams,
j) -- l).
\end{itemize}

\noindent The corrections (i)-(iv) have a 
          different relative impact:
\begin{itemize}
\item[-] Electroweak corrections of type (i) are typically of the order
  of a few percent (like in the Standard Model) and do not exhibit a
  strong dependence on any SUSY parameters.
\item[-] The main source of differences between the tree-level  
  and higher-order results are the corrections to the Higgs-boson
  self-energies (ii). They are responsible for changes in the physical
  masses $\Mh$ and $\MH$ and the effective mixing angle $\aeff$
  (via contributions to the renormalization constants, $\Zext$, for the
  external Higgs particles in the $S$-matrix elements, see
  \refse{subsec:zext})
  predicted for given values of $\tb$ and
  $\MA$.  At the \onel\ level these propagator corrections
  constitute the only source for the
  large correction of \order{\gf\mt^4}. At the \twol\ level they
  exclusively give rise to contributions of \order{\gf\als\mt^4} and
  of \order{\gf^2\mt^6}. In this sense the propagator corrections
  define a closed subset of diagrams, being responsible for a
  numerically large contribution.  
  %They also enter the renormalization
  %constants, $\Zext$ (see \refse{subsec:zext}) for the external Higgs
  %particles in the $S$-matrix elements. The latter are very important
  %since they constitute the link between the angle $\aeff$, which
  %enters the tree-level Higgs-boson vertices in the EPA method, and
  %the tree-level angle $\al$ of \refeq{higgsrotation}.
%
\item[-] Corrections to the final-state vertices (iii) are
  typically larger than those of type (i), but smaller than the
  Higgs-boson propagator corrections. At LEP2 energies they can reach at
  most 7-10\%~\cite{CPR} for very low or very large values of $\tb$,
  when the Yukawa couplings of the top or bottom quarks become strong.
\item[-] Finally, the box-diagram contributions (iv) depend strongly 
  on the center-of-mass energy. They are of the order of 2-3\% at LEP2
  energies and may reach 20\% for $\sqrt{s}=500$ GeV~\cite{higgsprod}.
\end{itemize}
It should be noted that initial-state QED corrections as well as
finite-width effects (allowing for off-shell decays of the Higgs and
the $Z$~boson) are not included
in our calculation. However, by incorporating our result into existing
codes, e.g.\ HZHA~\cite{hzha}, QED corrections and finite-width effects
can automatically be taken into account%
\footnote{
The implementation of our calculation into HZHA is currently
investigated~\cite{feynhiggsXSinHZHA}.%
}%
 .

%%%%%%%%%%%%%%%%%%%%%%%%%%%%%%%%%%%%%%%%%%%%%%%%%%%%%%%%%%%%%%
%%%%%%%%%%%%%%%%%%%%%%%%%%%%%%%%%%%%%%%%%%%%%%%%%%%%%%%%%%%%%%

\subsection{Higgs-boson masses and wave function renormalizations}
\label{subsec:zext}

Radiative corrections induce mixing between the $\cp$-even Higgs bosons,
even if their mass matrix has been diagonalized at the tree level%
\footnote{ 
We do not consider here possible $\cp$-violating mixing between neutral
Higgs bosons, which can occur if the MSSM Lagrangian contains complex
parameters~\cite{cphiggs}.  
}%
.  In the FD approach the higher-order corrected Higgs-boson masses,
denoted by $\Mh, \MH$, are derived by finding the poles of the
$h,H$-propagator matrix, which is given by
\BE
\hat{\bf D}_S = \ML \Delta_H    & \Delta_{hH} \\
    \Delta_{hH} & \Delta_h \MR
= i \ML q^2 - \mH^2 + \hSi_{HH}(q^2) & \hSi_{hH}(q^2) \\
\hSi_{hH}(q^2) & q^2 - \mh^2 + \hSi_{hh}(q^2) \MR^{-1} .
\label{eq:invprop}
\EE
Here the $\hSi(q^2)$ denote the renormalized Higgs-boson self-energies
(throughout the paper we denote tree-level mass
parameters by small letters and physical masses by capital
letters). For these self-energies we take the result of the complete
one-loop on-shell calculation of \citeres{CPRlet,CPR} together with
the dominant \twol\ correction of $\oaas$ obtained
in~\citeres{mhiggsletter,mhiggslong} and the sub-dominant corrections
of $\ogmzmts$ \cite{mhiggsRG1a,mhiggsRG1b,mhiggsRG2}.

Determining the poles of \refeq{eq:invprop} corresponds to solving the
equation
\BEA
\re \KKL  \KL q^2 - \mh^2 + \hSi_{hh}(q^2) \KR 
          \KL q^2 - \mH^2 + \hSi_{HH}(q^2) \KR 
        - \KL \hSi_{hH}(q^2) \KR^2 \KKR = 0 .
\label{eq:hmass}
\EEA

The wave function renormalization factors for the scalar Higgs bosons 
$\Hi$ are denoted as $\Zext_{\rS i}$. They are the finite residues of the $H$
and $h$ propagators, respectively,
\BEA
\Zext_{\rS 1}\equiv \Zext_{\rS H} &=& 
 \left.\left(1 + \re\hSipH(q^2) - \re\KL \frac{\hSihH^2(q^2)} {q^2
    - \mh^2 + \hSih(q^2)} \KR '\right)^{-1}\right|_{ q^2 = \MH^2}~, \non\\ 
\label{eq:zext}
\Zext_{\rS 2}\equiv \Zext_{\rS h} &=& 
 \left.\left(1 + \re\hSiph(q^2) - \re\KL \frac{\hSihH^2(q^2)}
            {q^2 - \mH^2 + \hSiH(q^2)} \KR '\right)^{-1}\right|_{q^2 = \Mh^2}~,
\EEA
where $\hSip(q^2) \equiv \frac{\partial}{\partial q^2} \hSi(q^2)$.

For a diagram with no mixing on the outgoing Higgs-boson line, $\Hi$,
the $S$-matrix element is given by the amputated on-shell Green's
function multiplied by the $ \KL \Zext_{\rS i} \KR ^{1/2}$.  In the
presence of mixing, i.e.\ $h \leftrightarrow H$ on the external line
(with the scalar $i$ being the final state particle) the respective
factor reads:
\BEA
\left.{ \frac{-\hSihH(q^2) \KL \Zext_{\rS i} \KR^{-1/2}
                              \KL q^2 - M^2_{\Hi} \KR}
      { \KL q^2 - \mH^2 + \hSiH(q^2) \KR 
        \KL q^2 - \mh^2 + \hSih(q^2) \KR 
      - \KL \hSihH(q^2) \KR ^2} }\right|_{q^2 = M^2_{\Hi}}~.
\EEA
Therefore, in this case, the amputated Green's function is effectively
multiplied by ($i'$ denotes the index of the ``supplementary'' Higgs
boson: $i' = 2(1)$ for $i = 1(2)$; formally $i' = 3 - i$)
\BE
 \KL \Zext_{\rS i} \KR ^{1/2}
 \frac{ -\hSihH \KL M^2_{\Hi} \KR} 
      { M^2_{\Hi}-m^2_{H_{i'}} +
        \hSi_{i'i'} \KL M^2_{\Hi} \KR }
\equiv 
\KL \Zext_{\rS i} \KR ^{1/2}\Zmix_{\rS i}~. 
\label{eq:zextm}
\EE

Exactly analogous equations hold for the pseudoscalar constants
$\Zext_{\rP i}$ and $\Zmix_{\rP i}$. Their effects, however, are
numerically much less important. Appropriate formulae for the
inclusion of the $\cZ$ factors into the higher-order corrected
vertices can be found in the Appendix.

%%%%%%%%%%%%%%%%%%%%%%%%%%%%%%%%%%%%%%%%%%%%%%%%%%%%%%%%%%%%%%
%%%%%%%%%%%%%%%%%%%%%%%%%%%%%%%%%%%%%%%%%%%%%%%%%%%%%%%%%%%%%%

\subsection{The $\aeff$ approximation}
\label{subsec:alpha}

The inclusion of the $\cZ$ factors on the external Higgs lines in the
on-shell calculation reproduces, in the approximation of the neglected
momentum dependence of the Higgs self-energies, 
the effect of using the higher-order corrected angle $\aeff$ in 
an improved Born approximation of the cross sections (see 
also \citere{hff}).

The dominant contributions for the Higgs-boson self-energies 
(of \order{\gf\mt^4} at the \onel\ level) are
obtained for $q^2=0$. Approximating the renormalized Higgs-boson
self-energies by
\BE
\hSi(q^2) \to \hSi(0) \equiv \hSi
\label{zeroexternalmomentum}
\EE
yields the Higgs-boson masses by re-diagonalizing the dressed mass
matrix
\BE
\label{deltaalpha}
\hat{\bf M}_{\rm Higgs}^{2} = \ML \mH^2 - \hSiH & -\hSihH \\ -\hSihH &
\mh^2 - \hSih \MR \stackrel{\De\al}{\longrightarrow}
   \ML \MH^2 & 0 \\ 0 &  \Mh^2 \MR ,
\EE
where $\Mh$ and $\MH$ are the corresponding higher-order-corrected
Higgs-boson masses.  In \citere{hff} it has been shown that in the
approximation with neglected external momentum the $\Zmix$ factors can
be written as follows in terms of $\Delta\alpha$, which is
the angle required for the re-diagonalization in \refeq{deltaalpha}:
\BEA
\Zmix_{\rS 1} \equiv \Zmix_{\rS H} &\stackrel{q^2 = 0}{\approx}&
 - \frac{\hSihH}{\MH^2 - \mh^2 + \hSih} = + \TDea, 
\label{ZHhTanDeltaalpha}
\\
\Zmix_{\rS 2} \equiv \Zmix_{\rS h} &\stackrel{q^2 = 0}{\approx}& 
- \frac{\hSihH}{\Mh^2 - \mH^2 + \hSiH} = - \TDea .
\label{ZhHTanDeltaalpha}
\EEA
It is important to note that, although it is not immediately visible,
both eqs.~(\ref{ZHhTanDeltaalpha},\ref{ZhHTanDeltaalpha}) yield the
same angle $\Delta\alpha$~\cite{hff}:
\BE
\Delta\alpha = \arctan 
               \frac{\Delta m^2 + \sqrt{(\Delta m^2)^2 + 4\,\hSihH^2}}
                    {-2\, \hSihH} ,
\EE
where
\BEA
\Delta m^2 \equiv (m_H^2 - \hSiH) -  (m_h^2 - \hSih) .
\EEA
The $\Zext$ factor can be expressed as 
\BEA
\Zext_{\rS 1} 
&\stackrel{q^2 = 0}{\approx}\Zext_{\rS 2} &\stackrel{q^2 = 0}{\approx}
{1\over 1 + \TQDea} = \CQDea.
\EEA
The consequences for
the couplings are demonstrated for $Zh$ production in the following
example for the $ZZh$ vertex.  The Born coupling $\tilde V_{ZZh} \sim
-\Samb$ is changed by the loop corrections, in terms of the $\cZ$
factors, according to
\BEA
\tilde V_{ZZh} &\sim& 
 \KL \Zext_{\rS h} \KR ^{\edz} \KKL -\Samb - \Zmix_{\rS h} \Camb \KKR \non\\
 &=& \CDea \KKL -\Samb - (-\TDea) \Camb \KKR \non \\
 &=& -\sin(\al - \be + \De\al) \non \\
 &=& -\sin(\aeff - \be) .
\label{eqn:zzhtilde}
\EEA
Analogous results hold for all Higgs vertices, including the $A\Hi$
vertices.  Therefore, the $\cZ$ factors effectively shift the
tree-level angle $\al$ by $\Delta\alpha$, yielding a loop-improved
angle
\BE
\aeff = \al + \De\al 
\label{def:alphaeff}
\EE
in this approximation.  

While the $\aeff$ approximation, i.e.\ using an improved Born result
for the cross sections
where the tree-level angle $\alpha$ is replaced by $\aeff$,
incorporates the dominant one-loop and two-loop contributions,
it is obvious from the discussion above that this approximation
neglects many effects included in a full FD calculation.
These are 
contributions from the full spectrum of the MSSM particles, the 
momentum dependence of the Higgs-boson self-energies, the gauge-boson 
and the fermion self-energy corrections, and in particular the
process-specific vertex and box corrections.

%%%%%%%%%%%%%%%%%%%%%%%%%%%%%%%%%%%%%%%%%%%%%%%%%%%%%%%%%%%%%%
%%%%%%%%%%%%%%%%%%%%%%%%%%%%%%%%%%%%%%%%%%%%%%%%%%%%%%%%%%%%%%

\subsection{Cross sections}
\label{subsec:crosssection}

In this section analytical formulae are presented for the cross
sections for the on-shell production of the Higgs bosons $e^+e^-\to
Z\Hi$, $e^+e^-\to A\Hi$ including the corrections (i)-(iii). Box
diagrams (iv) give another, more complicated, set of formfactors that
make the expressions quite lengthy and are hence are omitted here;
more details can be found in~\citere{boxes}.  However, we include the
box-diagram contributions, as described in~\citere{higgsprod}, in our
numerical programs~\cite{feynhiggs} and in the figures presented in
this paper.

The presented formalism for cross sections is general enough to
accommodate corrections of any order to 2- and 3-point vertex
functions.  Beyond the one-loop level, however, currently only \twol\
corrections to the scalar propagators have been
calculated~\cite{mhiggsletter,mhiggslong}.  Therefore, in the cross
section calculations we include all possible types of \onel\
corrections and the available \twol\ corrections to scalar
self-energies.  This is well justified because, as discussed above,
propagator corrections constitute a closed subset of the leading
\order{\gf\als\mt^4} and \order{\gf^2\mt^6} contributions. Therefore,
these \twol\ corrections are of particular relevance and interest.

\bigskip
The cross sections (in the CMS) for both processes (\ref{eetohZ}) and
(\ref{eetohA}) have the form:
\BEA
{d\sigma_{Z(A)H_i}\over d\Omega}=
{\lambda \KL s,M^2_{Z(A)},M^2_{\Hi} \KR \over 64\pi^2
s^2\left|D_Z(s)\right|^2}  \KL {\cal A}_1+{\cal
A}_2\cos^2\theta_{\rm CMS} \KR ,
\EEA
where $\lambda$ is the standard phase space factor,
\BEA
\lambda \KL s,m^2_1,m^2_2 \KR =
\sqrt{s^2+m_1^4+m_2^4-2sm_1^2-2sm_2^2-2m^2_1m^2_2} ,
\label{def:lambda}
\EEA
and ${\cal A}_1,{\cal A}_2$ are defined by 
\BEA
\label{eq:cr_sum_mat}
{\cal A}_1+{\cal A}_2\cos^2\theta_{\rm CMS} ={1\over
4}\sum_{pol} \KL {\cal M}{\cal M}^{*} \KR ,
\EEA
with ${\cal M}_{ZS}$ and ${\cal M}_{PS}$ as given below.
In the following, $\theta_{\rm CMS}$ denotes the scattering angle
$\theta_{\rm CMS} = < \hspace{-.45em} )\,(e^-, H_i)$ in the CMS.
The momenta of the incoming electron and positron are
denoted as $k_1$ and $k_2$, respectively. The momentum of the outgoing
$h,H$ is labeled with $q$, whereas the outgoing $Z,A$ momentum is
denoted as $p$, see \reffi{fig:diagrams}a.  The matrix elements for the
Higgs-strahlung process and the associated Higgs production read (in
the approximation of neglected box diagrams)
\BEA
{\cal M}_{ZS}^i=e\overline v(k_2)\ga^{\nu} \KKL \tilde V^{\mu\nu
i}_{ZZS} \KL \hat c_V-\hat c_A\ga^5 \KR + {\tilde V}^{\mu\nu i}_{\ga
ZS} {D_Z(s)\over D_{\ga}(s)} - \tilde V^{(0)\mu\nu i}_{ZZS}
{\hat\Sigma^T_{\ga Z}(s)\over D_{\ga}(s)} \KKR u(k_1)\epsilon_{\mu}(p) ,
\EEA
\BEA
{\cal M}_{PS}^{i}=e\overline v(k_2)\ga_{\mu} \KKL \tilde V^{\mu
ij}_{ZPS} \KL \hat c_V-\hat c_A\ga^5 \KR + {\tilde V}^{\mu ij}_{\ga
PS}{D_Z(s)\over D_{\ga}(s)} - \tilde V^{(0)\mu ij}_{ZPS}
{\hat\Sigma^T_{\ga Z}(s)\over D_{\ga}(s)} \KKR u(k_1) .
\EEA
For the corresponding expressions for the box contributions see
\citere{boxes}. 

In the above expressions $u(k_1)$ and $v(k_2)$ are spinors of the
incoming electron-positron pair, $\epsilon_{\mu}(p)$ is the
polarization vector of the outgoing $Z$. $\hat c_V, \hat c_A$ are the
renormalized vector and axial couplings of the $Z$ boson to an
electron positron pair, at the \onel\ level $\hat c_A=-1/4s_Wc_W+ \hat
c_A^{(1)}$, $\hat c_V= \KL -1+4s^2_W \KR /4s_Wc_W+ \hat
c_V^{(1)}$, $\cw^2 \equiv 1 - \sw^2 \equiv \MW^2/\MZ^2$.
$\hat\Sigma^T_Z(s)$, $\hat\Sigma_{\ga}(s)$ and
$\hat\Sigma^T_{\ga Z}(s)$ denote the renormalized photon and transverse
$Z$~boson self-energies.  $D_Z(s)$ and $D_{\ga}(s)$ are the inverse
$Z$ and photon propagators defined as
\BEA
D_Z(s) = s - M_Z^2 + \hat\Sigma^T_Z(s)~, \nonumber\\
D_{\ga}(s) = s + \hSi_\ga(s)~.
\EEA
Finally, $\tilde V$ denotes the effective neutral Higgs--gauge-boson
vertices with the one-loop form factors.  The explicit expression for
those vertices and for the matrix elements for Higgs-strahlung and
associated Higgs production can be found in the Appendix and in
\citere{CPR}.

%%%%%%%%%%%%%%%%%%%%%%%%%%%%%%%%%%%%%%%%%%%%%%%%%%%%%%%%%%%%%%
%%%%%%%%%%%%%%%%%%%%%%%%%%%%%%%%%%%%%%%%%%%%%%%%%%%%%%%%%%%%%%

\section{Numerical results}
\label{sec:num}

\subsection{Parameter choice}
\label{subsec:pars}

In the following we present numerical examples for the dependence of
the neutral Higgs-boson couplings and cross sections on $\tb$, $\Mh$,
and the mixing in the scalar top sector.  
In all plots, as a typical example, the set of
parameters listed in Table~\ref{tab:par} has been used, if not stated
differently.
\begin{table}[htb!]
\begin{center}
\begin{tabular}{|c|c|c|c|c|c|c|c|}
\hline
$\mt$ & $\mb$ &  $M_{\tilde{q}}$ & $M_{\tilde{l}}$ & $\mu$ & $M_2$ &
 $M_1$ & $M_3$ \\
\hline
&&&&&&&\\
174.3 & 4.5 & 1000 & 300 & 200 & 200 & $\frac{5}{3}\sw^2/\cw^2 M_2$ & 
$\frac{\als}{\al_{\rm em}}\sw^2 M_2$\\
&&&&&&&\\
\hline
\end{tabular}
\caption{Quark masses and SUSY parameters (in GeV) used in the 
numerical analysis.
\label{tab:par}} 
\end{center}
\end{table}
$m_t$ and $m_b$ in Table~\ref{tab:par} are the quark pole masses.
$M_{\tilde{q},\tilde{l}}$ denote the soft SUSY-breaking parameters in
the scalar quark and lepton sector, respectively (in the following we
also use the notation $\msusy\equiv M_{\tilde{q}}$) and $M_3
\equiv \mgl$ denotes the gluino mass.  The mixing in the scalar top
sector, which plays a prominent role in the physics of the MSSM Higgs
sector, is controlled by the off-diagonal term in the scalar-top mass
matrix, $\mt \Xt \equiv \mt (\At - \mu\cot\beta)$, in the convention
of~\citere{mhiggslong}. In our analysis we have focused on two
different values of $\Xt$ leading to two extreme values of the physical
Higgs-boson mass $\Mh$, as suggested by
\citeres{mhiggslong,mhiggslle,tbexcl,benchmark}.  The lightest MSSM
Higgs-boson mass as a function of $\Xt/\msusy$ has a minimum at
$\Xt/\msusy \approx 0$, denoted in the following as ``no mixing''
case.  A maximum value is reached at $\Xt/\msusy \approx 2$, denoted
further as ``maximal mixing''.  For the sbottom sector we assume a
universal trilinear coupling, $\Ab = \At$. These values and the
parameters in Table~\ref{tab:par} are understood to be input
parameters for the diagrammatic calculation in the on-shell
renormalization scheme. 

Below we will also perform comparisons with
results obtained in the framework of the RG improved one-loop
EPA, where the
input parameters are understood as \msbar\ quantities.
To ensure consistency, in the latter case we have transformed
the on-shell SUSY input parameters into the corresponding
\msbar\ values as discussed in \citere{bse}.  
The results shown below for the higher-order corrected Higgs-boson 
masses and the mixing angle within the RG improved one-loop EPA have been 
obtained with the Fortran program {\em subhpole} 
(based on \citeres{mhiggsRG1a,mhiggsRG1b,bse}).

%%%%%%%%%%%%%%%%%%%%%%%%%%%%%%%%%%%%%%%%%%%%%%%%%%%%%%%%%%%%%%
%%%%%%%%%%%%%%%%%%%%%%%%%%%%%%%%%%%%%%%%%%%%%%%%%%%%%%%%%%%%%%

\subsection{2-loop corrections to masses and effective couplings}

\begin{figure}[htb!]
        \begin{center}
\begin{tabular}{p{0.48\linewidth}p{0.48\linewidth}}
\mbox{\epsfig{file=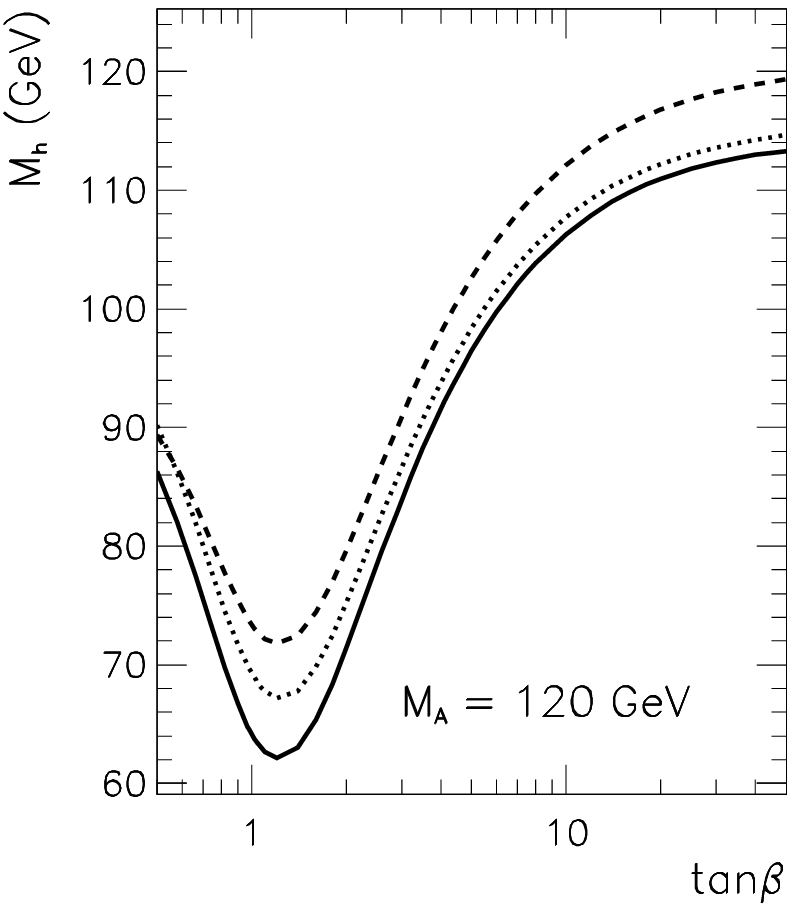,width=\linewidth,height=0.9\linewidth}}&
\mbox{\epsfig{file=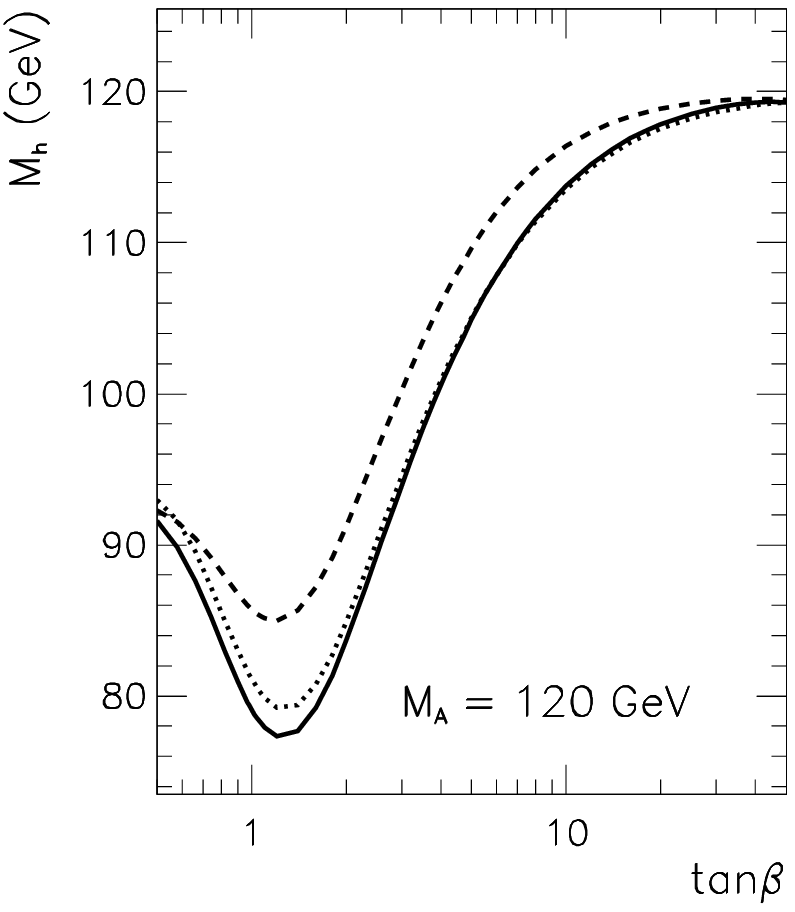,width=\linewidth,height=0.9\linewidth}}\\
\end{tabular}
\vskip -5mm 
\caption{$M_h$ as a function of $\tb$ for $M_A=120$ GeV and the parameters
of Table~\ref{tab:par}.  The no-mixing (maximal-mixing) case is shown
in the left (right) plot.  The solid line represents the \twol\ FD result,
the dotted line shows the RG improved one-loop EPA result and the dashed 
line shows the \onel\ FD result.}
\label{fig:mh_tb120}
\end{center}
\end{figure}

The dependence of the physical neutral Higgs-boson masses on the MSSM
parameters at the 2-loop level has been extensively discussed in the
literature~\cite{mhiggsRG1a,mhiggsRG1b,mhiggsRG2,mhiggsEP1,mhiggsEP2,maulpaul,mhiggsletter,mhiggslong,mhiggslle}.
As an illustration, we present in Fig.~\ref{fig:mh_tb120} the
dependence of $\Mh$ on $\tb$ for a relatively low $\MA$ value, $\MA =
120 \gev$. The \twol\ FD result is compared with the RG improved one-loop
EPA and also with the \onel\ FD result in the no-mixing and in the
maximal-mixing scenario in the left and in the right plot of
\reffi{fig:mh_tb120}, respectively.  In both scenarios $\Mh$ shows a
similar behavior: a minimum is reached around $\tb \approx 1$, maximum
values are reached for the largest $\tb$ values%
\footnote{One should keep in mind, however, that for fixed 
$\mt = 174.3 \gev$ and $\msusy \lsim 1$ TeV $\tb$ around~1 is already
excluded~\cite{tbexclexp,tbexcl} via Higgs-boson searches.
}%
. In the no-mixing scenario the FD result is always smaller than the
RG improved one-loop EPA value for $\Mh$, with a maximum difference around 
$\tb = 1$ of up to $5 \gev$. In the maximal-mixing scenario both result 
mostly agree. Note, however, that this behavior changes for larger values of
$\MA$, where the maximum value of $\Mh$ obtained in the FD approach is
a few GeV larger than the corresponding RG improved one-loop EPA
value~\cite{mhiggslong,mhiggslle,bse}.

%%%%%%%%%%%% F I G U R E %%%%%%%%%%%%%%%%%%%%%%%%%%%%%%%%%%%%%
%
%
%%%%%%%%%%%% F I G U R E %%%%%%%%%%%%%%%%%%%%%%%%%%%%%%%%%%%%%

Approximate results for the cross sections have often been obtained in
the literature on the basis of improved Born results, where the
effective mixing angle $\aeff$, see \refeq{def:alphaeff}, 
and the higher-order corrected Higgs-boson
masses $\Mh$ and $\MH$ have been evaluated within the RG improved one-loop
EPA. We will in the following compare our FD result for the cross sections
with this approximation, to which we will refer as ``RG $\aeff$
approximation''. As mentioned above, the results in the RG $\aeff$
approximation have been obtained using the program {\em subhpole}
(based on \citeres{mhiggsRG1a,mhiggsRG1b,bse}). 

In order to disentangle the effect of different contributions in this
comparison, we first discuss the results for the effective
$Z$--Higgs-boson couplings in the two approaches. In
\refse{subsec:alpha} we have shown that the contribution of the wave
function renormalization factors of the Higgs bosons is given
by the effective mixing angle $\aeff$ evaluated in the FD approach, if
the momentum dependence of the Higgs-boson self-energies is neglected.
However, in the actual cross-section calculation in the FD approach the
momentum dependence is included at the available (currently \onel) level.
Therefore, for a better comparison of
the quantities actually entering the cross section calculation in the
two approaches, we formally define $\aeff$ in the FD approach as
(in analogy to \refeq{ZhHTanDeltaalpha})
\BEA
\aeff^h = - \arctan \Zmix_{\rS h} ,
\label{eq:aeffh}
\EEA
where the $\Zmix_{\rS h}$ is given by the exact
expression~(\ref{eq:zextm}) (with Higgs self-energies calculated at
$q^2=M_h^2$), not by the $q^2=0$ approximation of
eq.~(\ref{ZhHTanDeltaalpha})%
\footnote{ 
It should be noted that $\aeff^H$
defined as $\aeff^H = \arctan\Zmix_{\rS H}$ slightly differs
from $\aeff^h$.%
}%
.

\begin{figure}[ht!]
\begin{center}
\begin{tabular}{p{0.48\linewidth}p{0.48\linewidth}}
\mbox{\epsfig{file=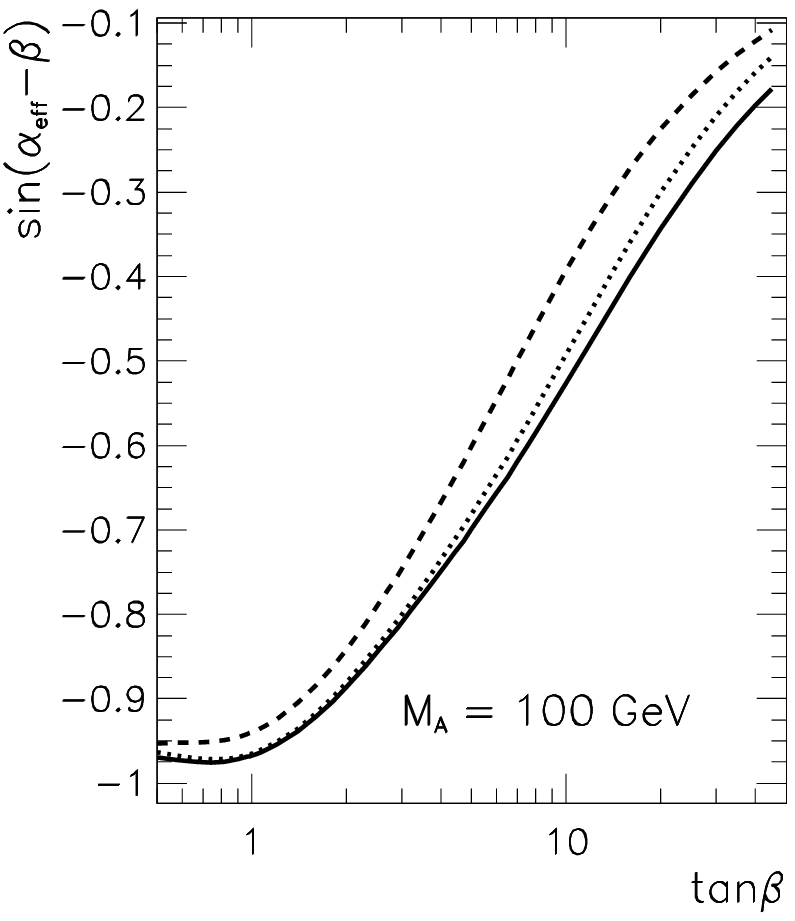,width=\linewidth,height=0.78\linewidth}}&
\mbox{\epsfig{file=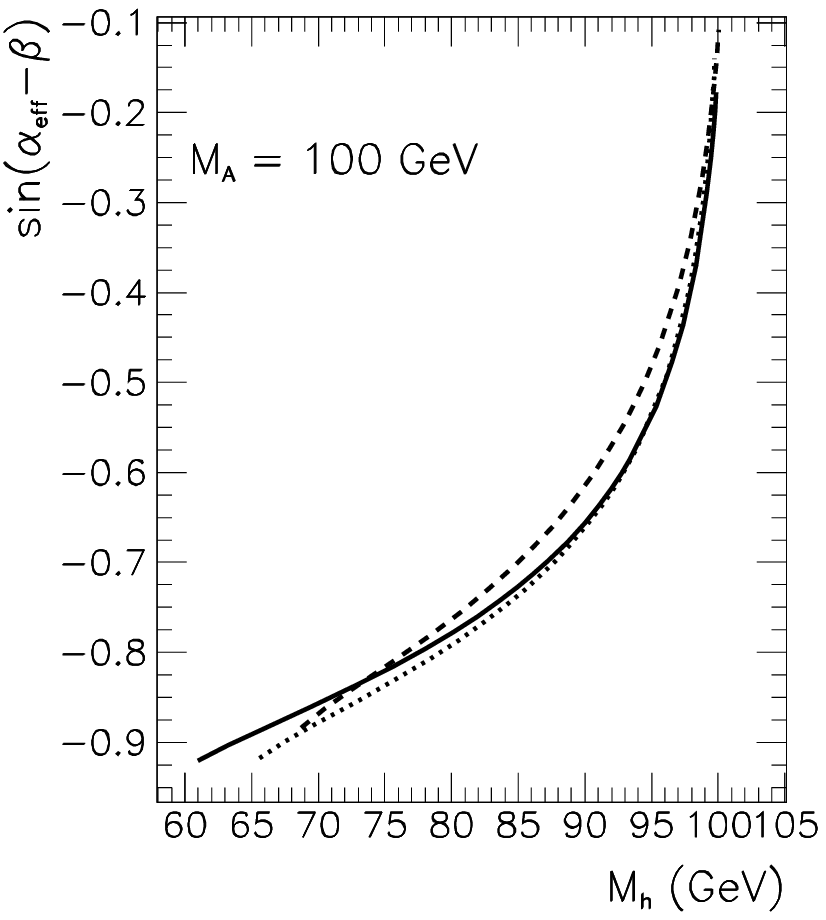,width=\linewidth,height=0.78\linewidth}}\\
\mbox{\epsfig{file=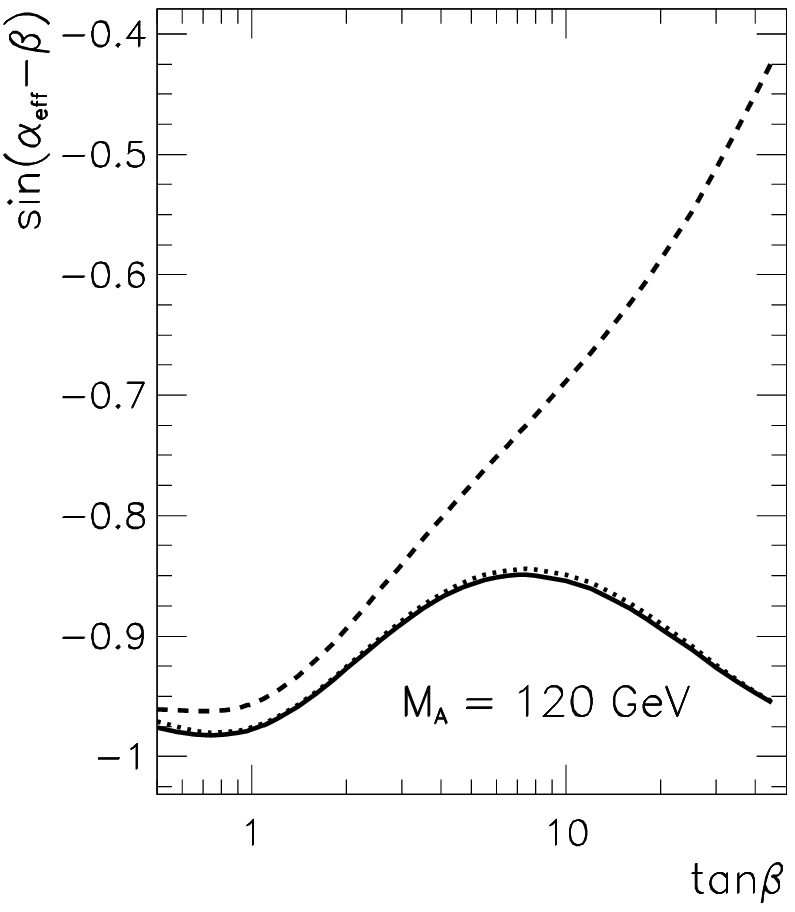,width=\linewidth,height=0.78\linewidth}}&
\mbox{\epsfig{file=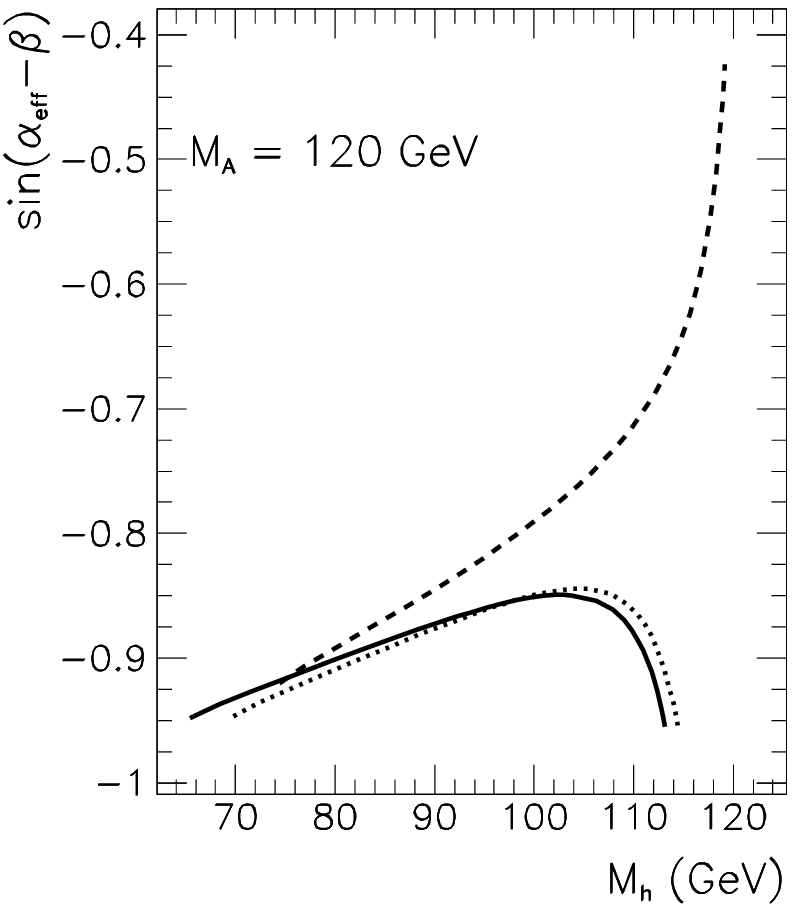,width=\linewidth,height=0.78\linewidth}}\\
\mbox{\epsfig{file=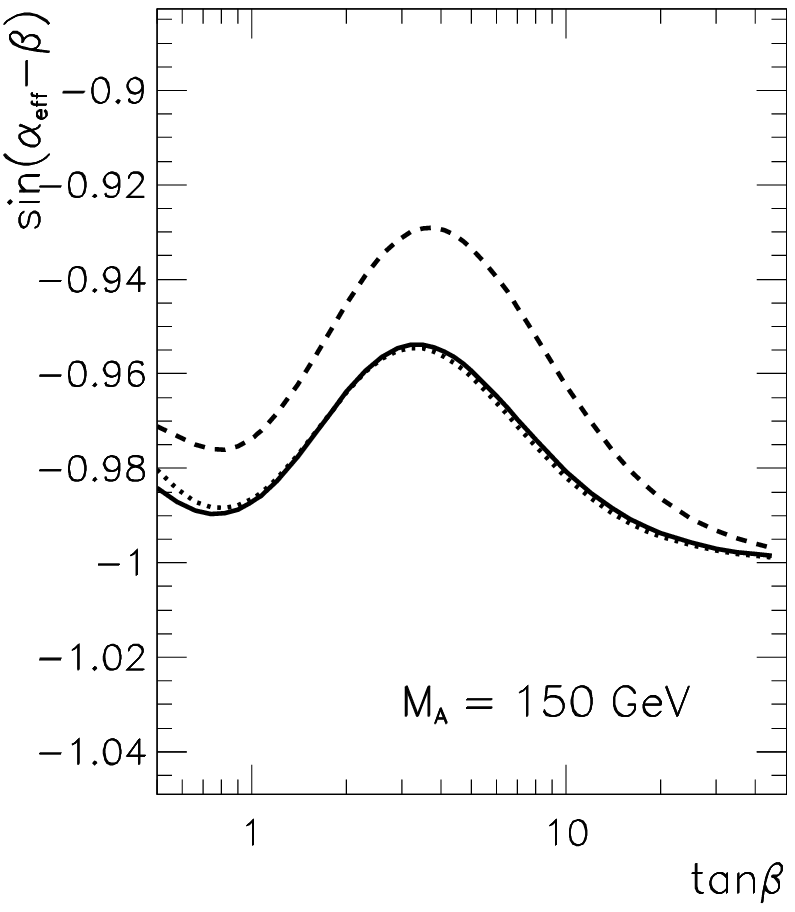,width=\linewidth,height=0.78\linewidth}}&
\mbox{\epsfig{file=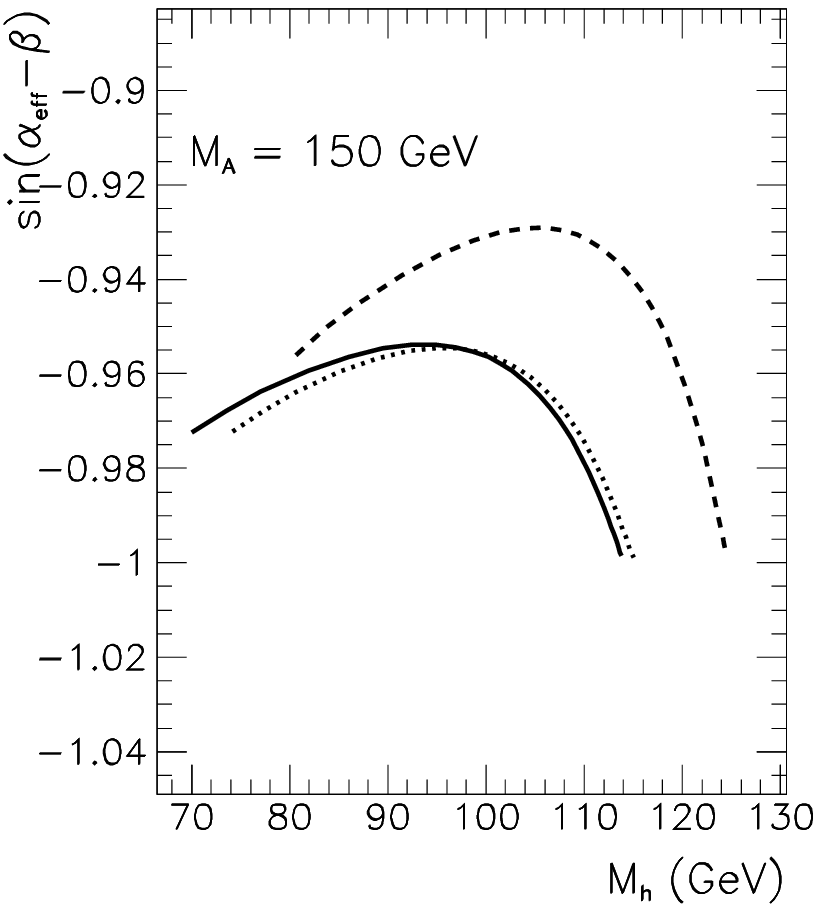,width=\linewidth,height=0.78\linewidth}}\\
\end{tabular}
\caption{$\sin(\aeff - \be)$ as a function of $\tb$ (left plots) and
$\Mh$ (right plots) for $\Xt=0$ and the parameters of Table~\ref{tab:par}.
The solid line represents the \twol\ FD result (see \refeq{eq:aeffh}),
the dotted line shows the RG improved one-loop EPA result, and the dashed 
line shows the \onel\ FD result.}
\label{fig:sba}
\end{center}
\end{figure}

Using this definition, in Fig.~\ref{fig:sba} the dependence of 
$\sin(\aeff - \be)$ on $\tb$
and $\Mh$ is shown in both approaches, for $\MA = 100, 120, 150 \gev$
and the no-mixing scenario, $\Xt=0$. 
Since $\Mh$ is a derived quantity and not an input parameter in our
approach, the parameter that is actually varied in the plots shown as
function of $\Mh$ is $\tb$. For simplicity, i.e.\ in order to avoid a
non-functional behavior, in all plots shown as function of $\Mh$ in this 
paper we restrict the $\tb$ region to $\tb>1.6$ (as mentioned above, 
for $\mt = 174.3 \gev$ and $\msusy \lsim 1$ TeV $\tb$ values around~1 are
already excluded~\cite{tbexclexp,tbexcl} via Higgs-boson searches.)

As can be seen in Fig.~\ref{fig:sba}, the agreement between the FD
two-loop result for $\aeff^h$ and the result within the RG improved one-loop
EPA is in general remarkably good (while large deviations can appear
compared to the FD one-loop result, see e.g.\ the middle plots in
Fig.~\ref{fig:sba}). For most of the $\tb$ range the FD two-loop result
and the result within the RG improved one-loop EPA differ
by not more than 5\%,
larger deviations can be observed only for $\MA = 100 \gev$ and
$\tb>10$ (left upper plot in Fig.~\ref{fig:sba}), where 
$\sin(\aeff - \be)$ itself is
small. Even in this case the $\sin(\aeff - \be)$ values in both approaches 
agree very well with each other when expressed in terms of the physical 
Higgs mass $\Mh$.
For the maximal-mixing case, $\Xt/\msusy=2$, the differences between the
effective couplings obtained in both methods are even smaller.

It should be noted that the behavior of $\sin(\aeff - \be)$ in the limit
of large $\tan\beta$ is quite different for small and large
pseudoscalar masses. This behavior changes for $\MA$ between 100 and
150 GeV; the actual value depends on the stop mixing parameter
$\Xt$ (see also Fig.~9 in \citere{hff}.)

%%%%%%%%%%%% F I G U R E %%%%%%%%%%%%%%%%%%%%%%%%%%%%%%%%%%%%%
%
%
%%%%%%%%%%%% F I G U R E %%%%%%%%%%%%%%%%%%%%%%%%%%%%%%%%%%%%%

%%%%%%%%%%%%%%%%%%%%%%%%%%%%%%%%%%%%%%%%%%%%%%%%%%%%%%%%%%%%%%
%%%%%%%%%%%%%%%%%%%%%%%%%%%%%%%%%%%%%%%%%%%%%%%%%%%%%%%%%%%%%%

\subsection{Results for the cross sections}

Differences in the Higgs-production cross sections 
between our FD result (containing the complete one-loop result and the
dominant two-loop corrections) and the RG $\aeff$ approximation
have a two-fold origin: the different predictions for
the values of $\Mh$ and $\aeff$, which we compared in the previous
section, and the additional contributions contained in the FD result
(i.e.\ the one-loop 3- and 4-point vertex functions and the 2-point
contributions that are not contained in $\aeff$).

\begin{figure}[htb!]
\begin{center}
\begin{tabular}{p{0.48\linewidth}p{0.48\linewidth}}
\mbox{\epsfig{file=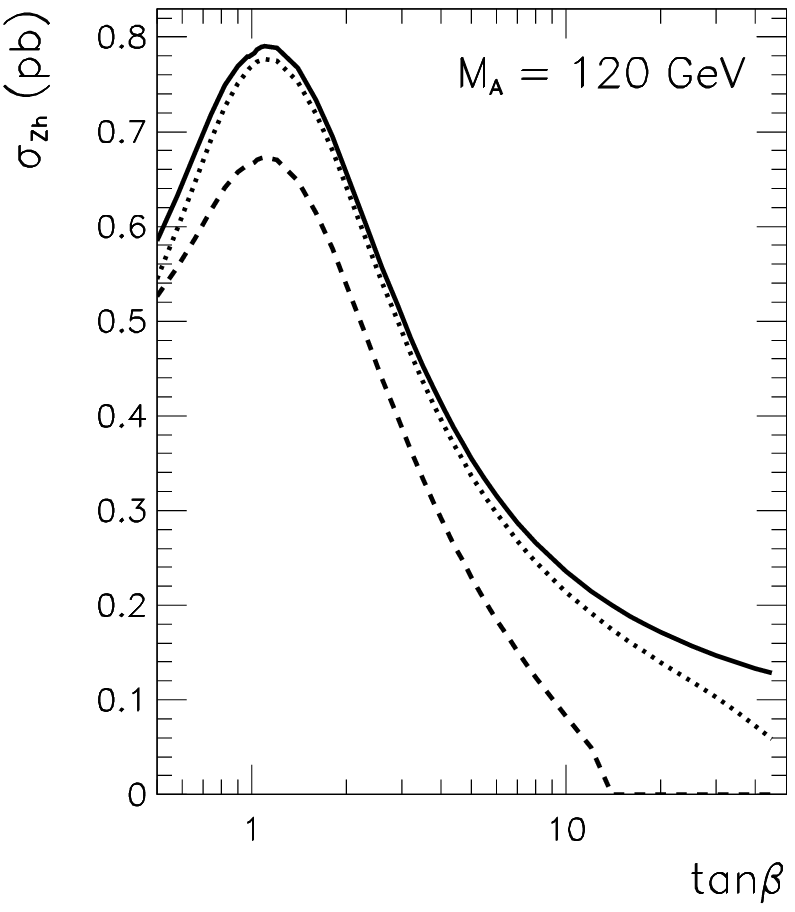,width=\linewidth,height=0.9\linewidth}}&
\mbox{\epsfig{file=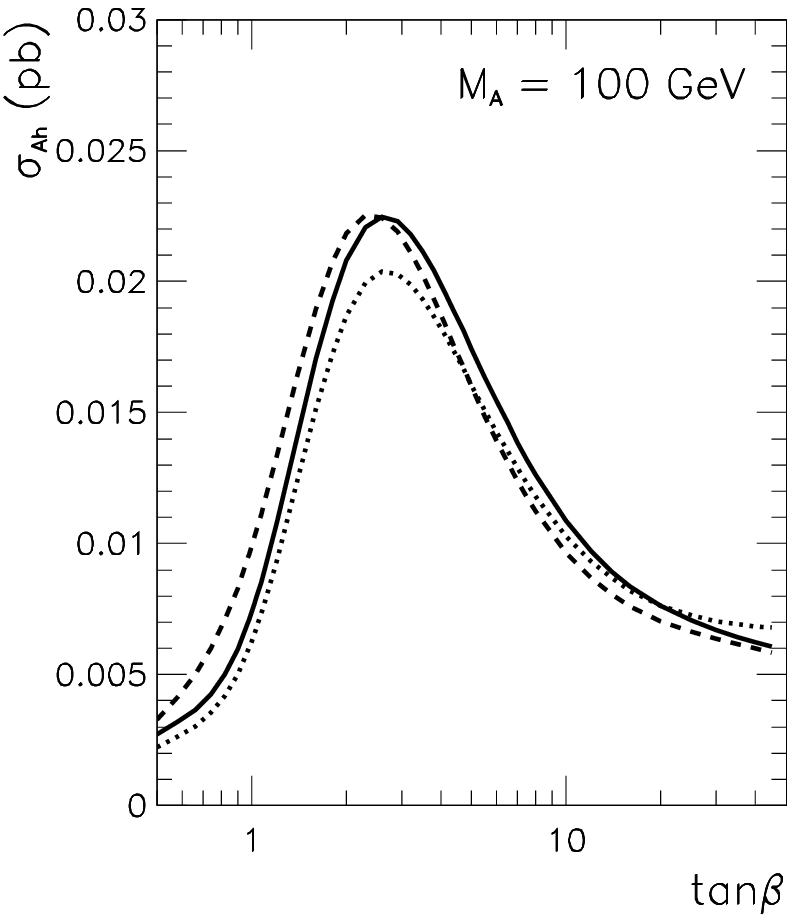,width=\linewidth,height=0.9\linewidth}}\\
\mbox{\epsfig{file=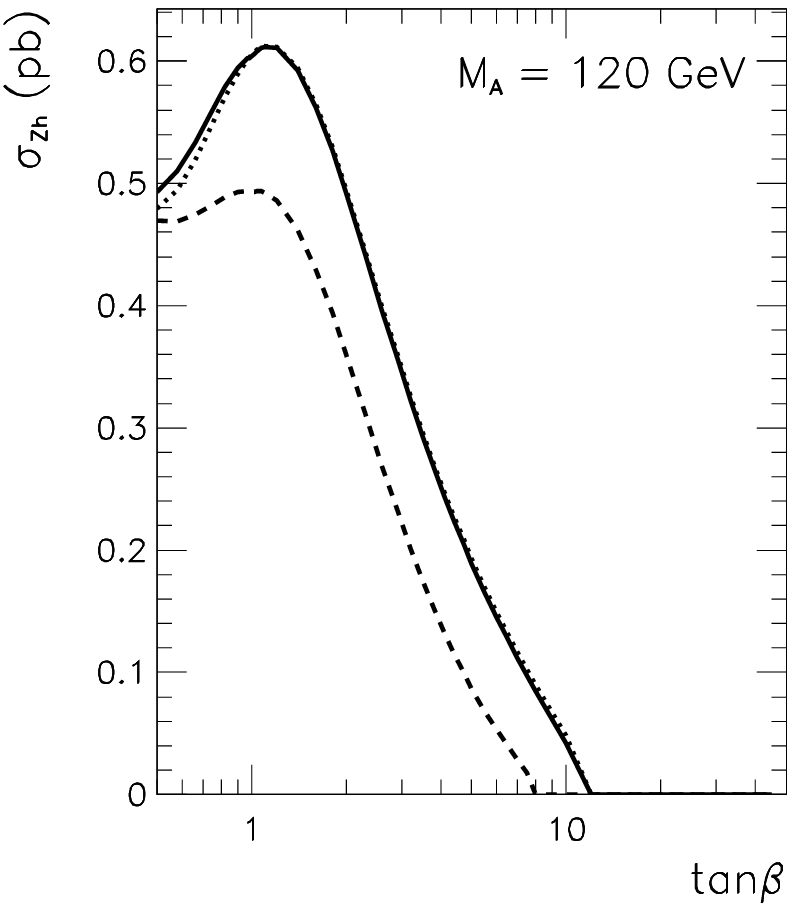,width=\linewidth,height=0.9\linewidth}}&
\mbox{\epsfig{file=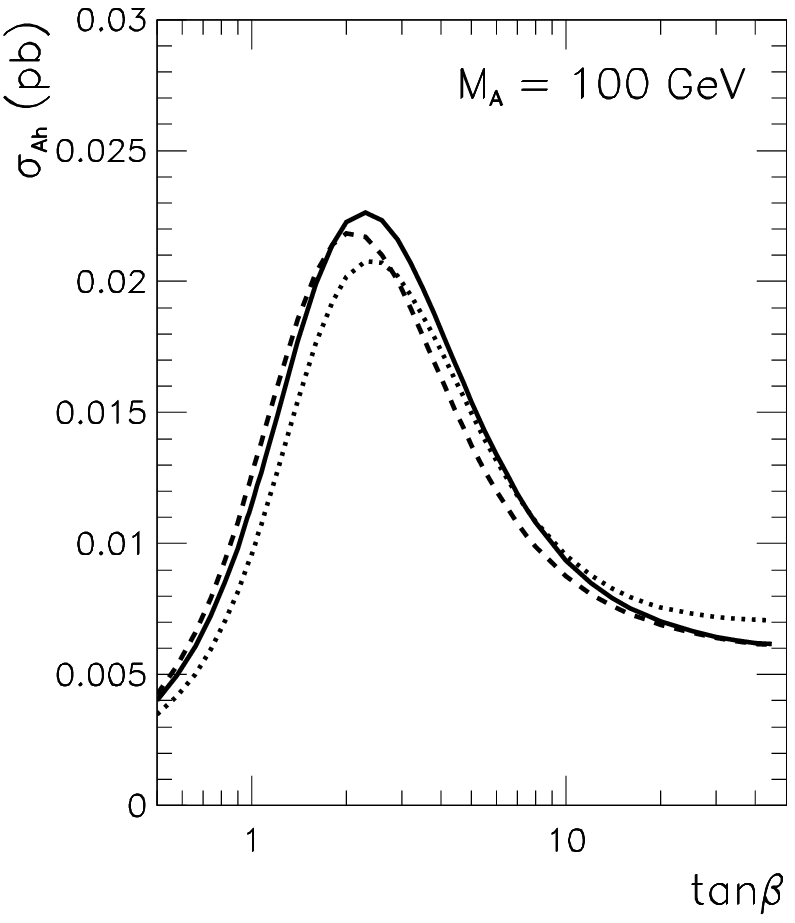,width=\linewidth,height=0.9\linewidth}}\\
\end{tabular}
\vskip -5mm 
\caption{$\sigma_{Zh}$ and $\sigma_{Ah}$ at $\sqrt{s} = 206 \gev$ 
as a function of $\tb$ for two values of $\MA$ and the parameters of
Table~\ref{tab:par}.  The upper (lower) row contains the result for
the no- (maximal)-mixing scenario.  The solid line represents the \twol\
FD result, the dotted line shows the result for the RG $\aeff$
approximation and the dashed line shows the \onel\ FD result.}
\label{fig:sigma_tb_s206}
\end{center}
\end{figure}

\begin{figure}[htb!]
\begin{center}
\begin{tabular}{p{0.48\linewidth}p{0.48\linewidth}}
\mbox{\epsfig{file=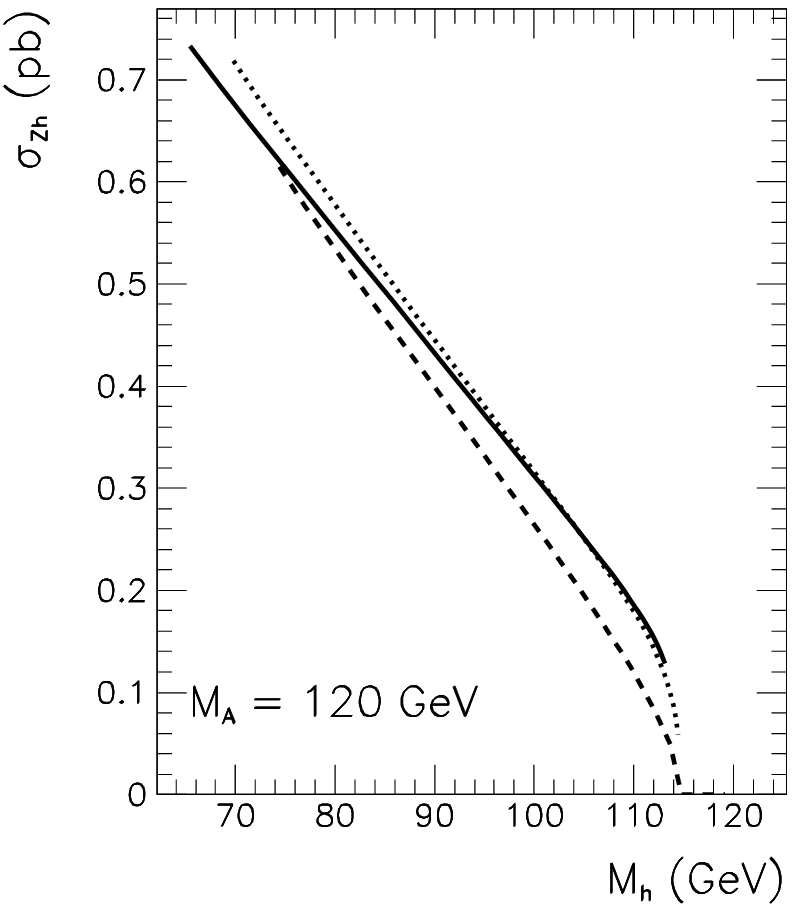,width=\linewidth,height=0.9\linewidth}}&
\mbox{\epsfig{file=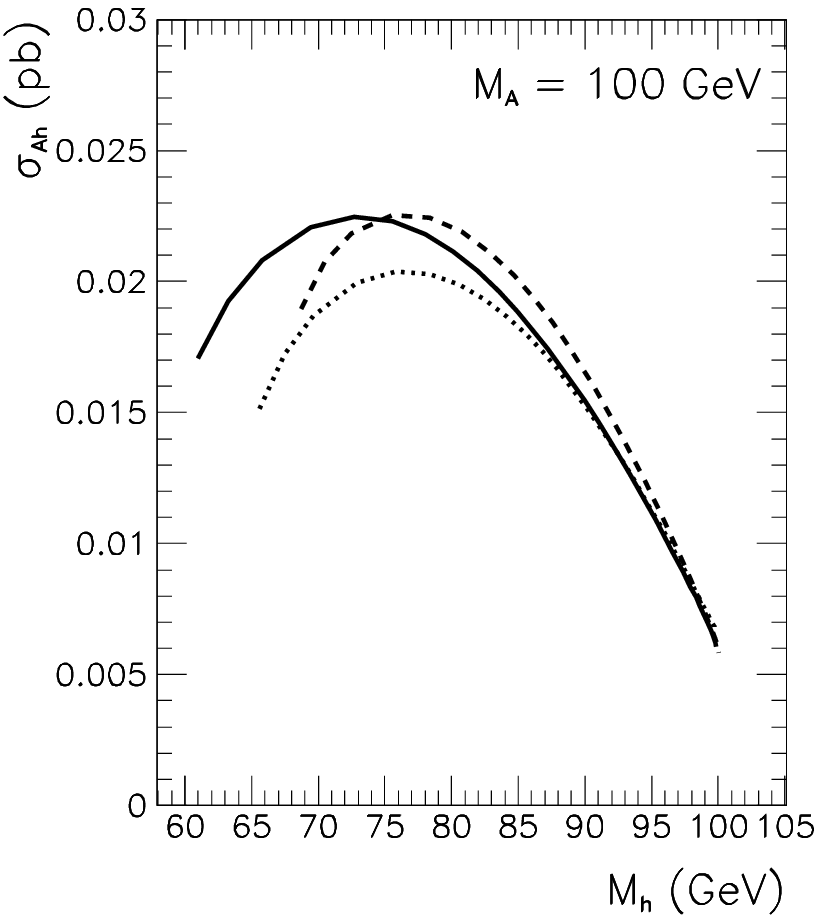,width=\linewidth,height=0.9\linewidth}}\\
\mbox{\epsfig{file=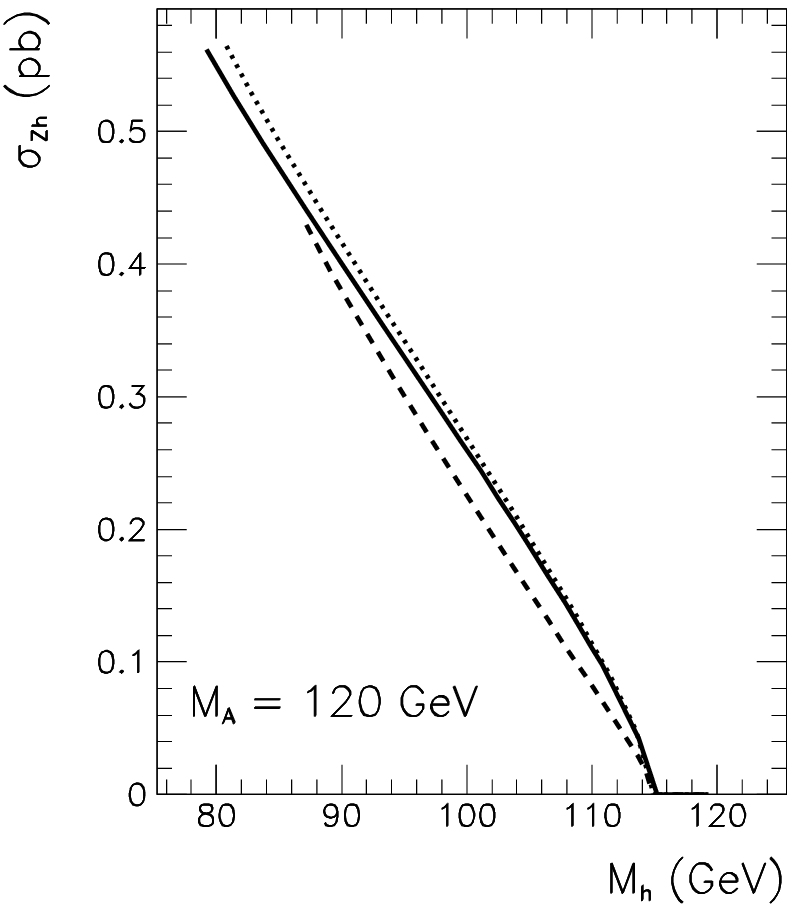,width=\linewidth,height=0.9\linewidth}}&
\mbox{\epsfig{file=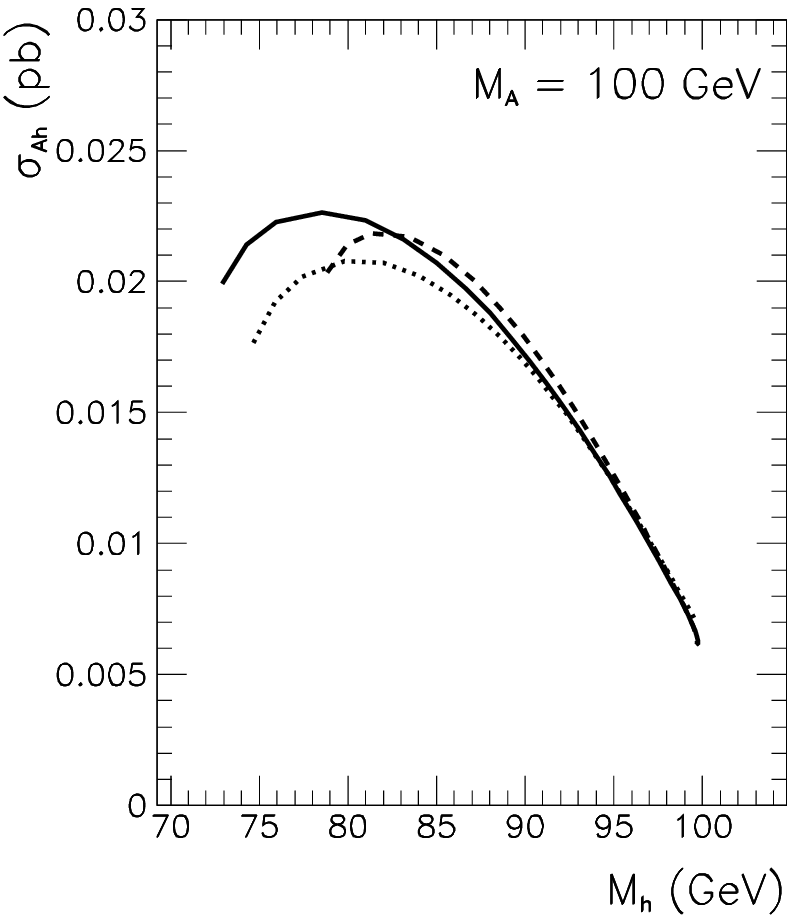,width=\linewidth,height=0.9\linewidth}}\\
\end{tabular}
\vskip -5mm 
\caption{$\sigma_{Zh}$ and $\sigma_{Ah}$ at $\sqrt{s} = 206 \gev$ 
as a function of $\Mh$ for two values of $\MA$ and the parameters of
Table~\ref{tab:par}.  The upper (lower) row contains the result for
the no- (maximal)-mixing scenario.  The solid line represents the \twol\
FD result, the dotted line shows the result for the RG $\aeff$
approximation and the dashed line shows the \onel\ FD result.}
\label{fig:sigma_mh_s206}
\end{center}
\end{figure}

In \reffi{fig:sigma_tb_s206} we present the cross sections for the two
production channels for a LEP2 energy of $\sqrt{s}=206$ GeV, in the
no-mixing and the maximal-mixing scenario, as a function of
$\tb$. \reffi{fig:sigma_mh_s206} shows the same results as a function
of $\Mh$. 
At LEP2 energies, the box diagram contributions are small,
of the order of 2-3\%~\cite{boxes,higgsprod}, and do not modify the
cross section behavior in a significant way.

%%%%%%%%%%%% F I G U R E %%%%%%%%%%%%%%%%%%%%%%%%%%%%%%%%%%%%%
%%%%%%%%%%%% F I G U R E %%%%%%%%%%%%%%%%%%%%%%%%%%%%%%%%%%%%%

%%%%%%%%%%%% F I G U R E %%%%%%%%%%%%%%%%%%%%%%%%%%%%%%%%%%%%%
%%%%%%%%%%%% F I G U R E %%%%%%%%%%%%%%%%%%%%%%%%%%%%%%%%%%%%%

As can be seen from both figures, the cross sections for the
Higgs-strahlung process corresponding to the FD result and the RG
$\aeff$ approximation are close to each other,
with differences of the order of a few per cent.  The only exception
occurs at $X_t\approx 0$ and large $\tb$. In this case the difference can 
amount up to 50\% when the cross sections are expressed as a function of 
$\tb$, but is considerably reduced
when they are calculated in terms of $\Mh$.
The \twol\ result in the FD
calculation is always above the \onel\ result; the difference can be
sizable, but is then mostly due to the kinematical effect that the
values for $\Mh$ at \onel\ are much larger than at the \twol\
level. This effect is especially pronounced at the kinematical
endpoints, i.e.\ for large $\tb$.

Associated $Ah$ production is of interest at LEP2 energies only for 
sufficiently low $\MA \leq 120$ GeV, otherwise it is kinematically
forbidden. Therefore we restricted our plots for $\sigma_{Ah}$
in~\reffis{fig:sigma_tb_s206} and~\ref{fig:sigma_mh_s206} to $\MA =
100 \gev$.  The difference between the FD result and the RG $\aeff$
approximation is larger than for
$\sigma_{Zh}$ and may reach about 20\% for not too large $\tb$ values.
The result of the \twol\ FD calculation is mostly above the \onel\
result, but the differences are much smaller than for the
Higgs-strahlung process. 

For both processes, the differences between the FD result and the 
RG $\aeff$ approximation are
not particularly pronounced, however visibly larger than those induced
by the modifications of $\aeff$ only. For LEP2 energies they can be
attributed mostly to the vertex corrections to the $ZZh$ and $ZhA$
couplings~\cite{CPR}.

%%%%%%%%%%%% F I G U R E %%%%%%%%%%%%%%%%%%%%%%%%%%%%%%%%%%%%%
\begin{figure}[htb!]
\begin{center}
\begin{tabular}{p{0.48\linewidth}p{0.48\linewidth}}
\mbox{\epsfig{file=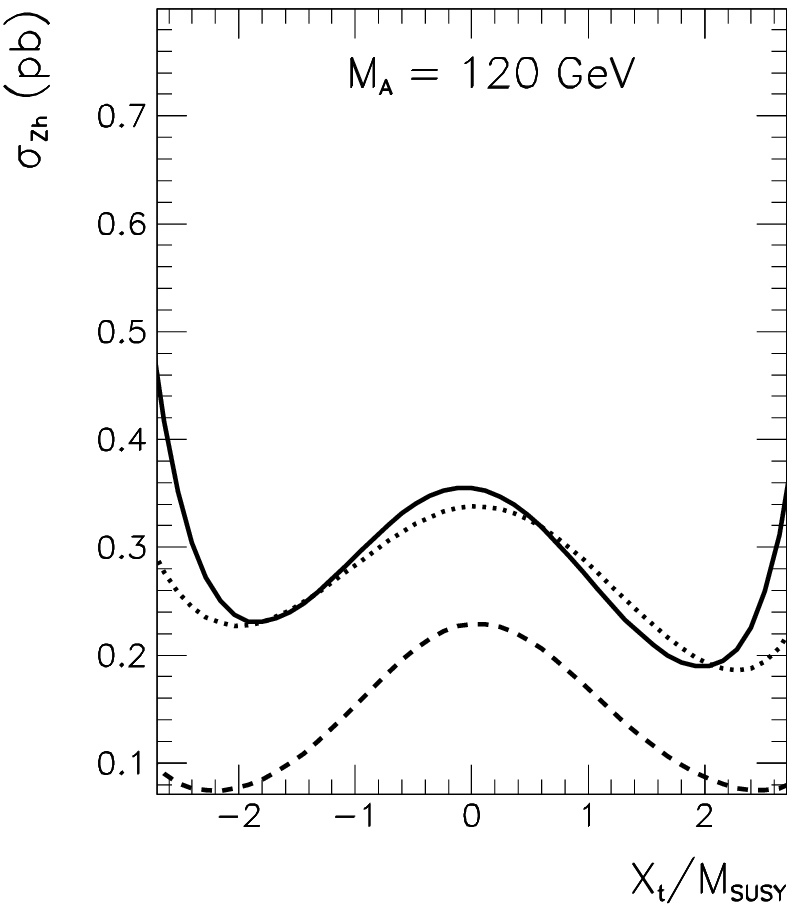,width=\linewidth,height=0.9\linewidth}}&
\mbox{\epsfig{file=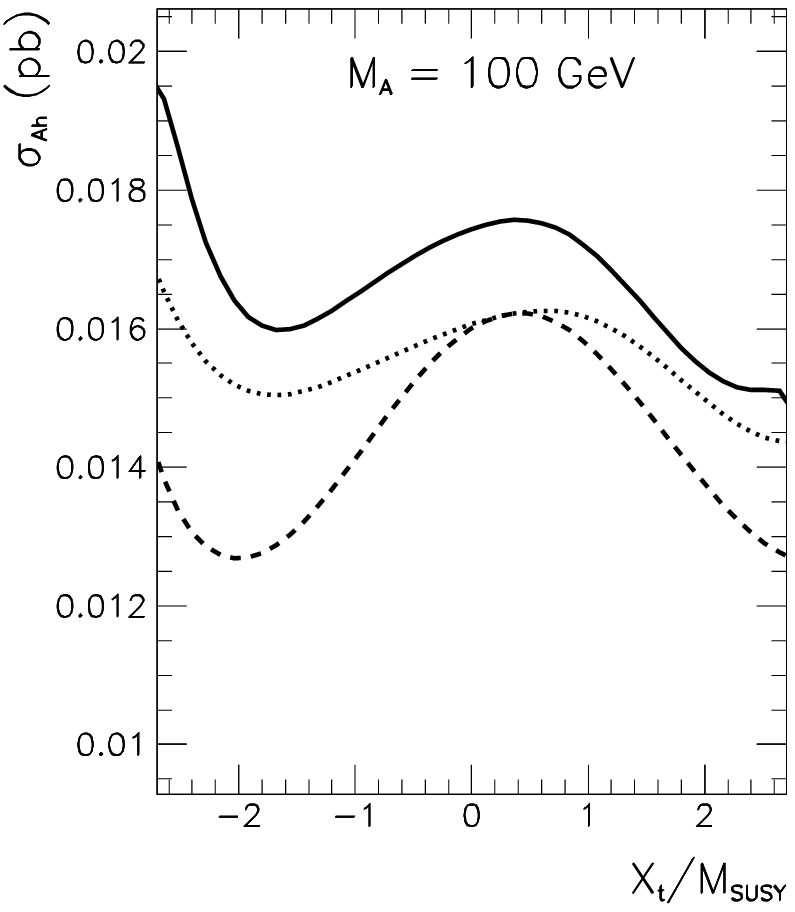,width=\linewidth,height=0.9\linewidth}}\\
\end{tabular}
\vskip -5mm 
\caption{$\sigma_{Zh}$ and $\sigma_{Ah}$ as a function of 
$\Xt/\msusy$ for $\sqrt{s} = 206 \gev$, $\tb = 5$ and the parameters of
Table~\ref{tab:par}. The solid line represents the \twol\ FD result,
the dotted line shows the RG $\aeff$ approximation and the dashed line shows
the \onel\ FD result.}
\label{fig:sigma_xt_s206}
\end{center}
\end{figure}
%%%%%%%%%%%% F I G U R E %%%%%%%%%%%%%%%%%%%%%%%%%%%%%%%%%%%%%

The dependence on the $\Stop$ mixing is depicted in
\reffi{fig:sigma_xt_s206}.  The main effect on $\sigma_{Zh},
\sigma_{Ah}$ is kinematical and arises from the change of $\Mh$ with 
$\Xt/\msusy$. This effect leads to a visible (additional) asymmetry in
the \twol\ FD results, whereas the RG $\aeff$ approximation and the
\onel\ FD result show a
weaker asymmetry in $\Xt/\msusy$.  As can be seen from the figure, the
differences between the methods are much larger (typically more then
10\%) when associated production is considered, with the above-mentioned 
``kinematical'' asymmetry further increased by the inclusion
of the vertex corrections.

%%%%%%%%%%%% F I G U R E %%%%%%%%%%%%%%%%%%%%%%%%%%%%%%%%%%%%%
\begin{figure}[htb!]
\begin{center}
\begin{tabular}{p{0.48\linewidth}p{0.48\linewidth}}
\mbox{\epsfig{file=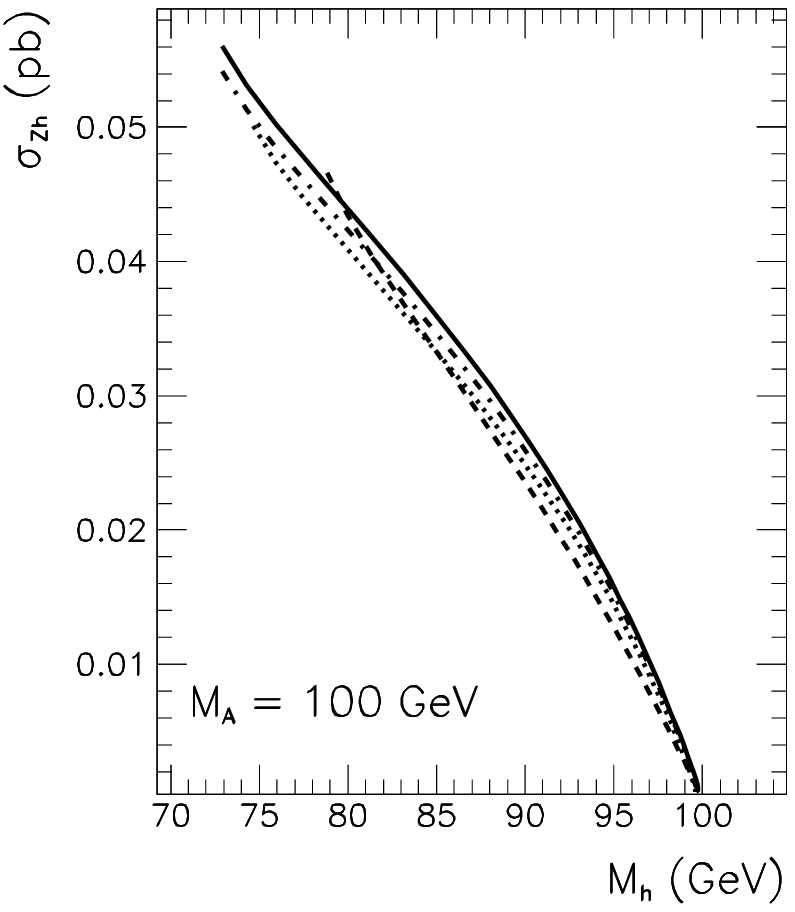,width=\linewidth,height=0.9\linewidth}}&
\mbox{\epsfig{file=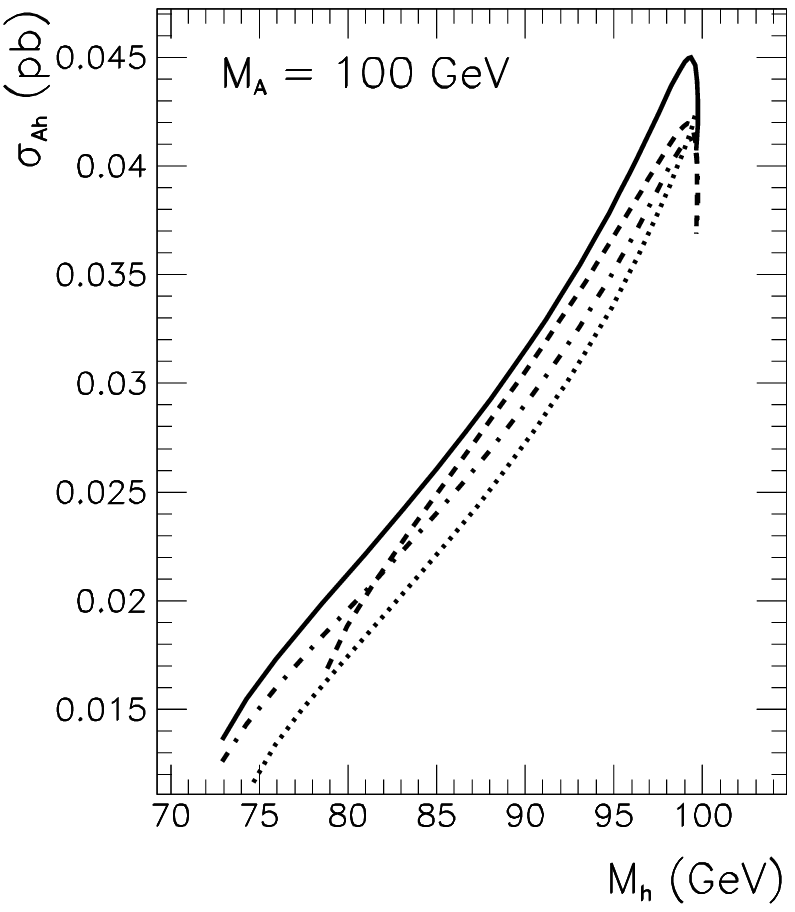,width=\linewidth,height=0.9\linewidth}}\\
\mbox{\epsfig{file=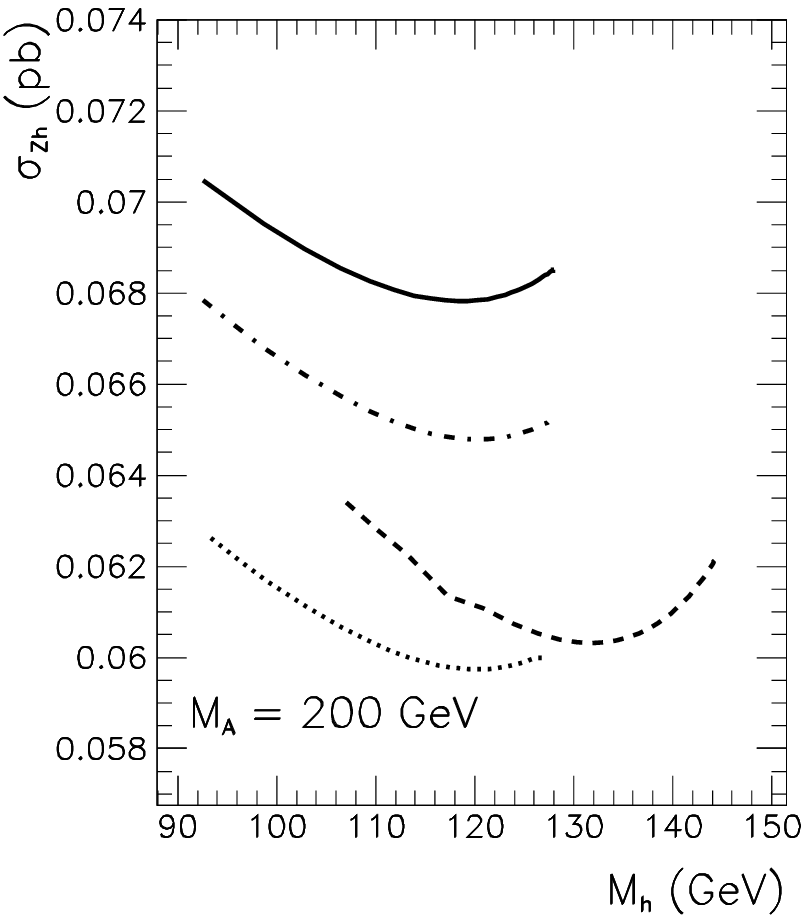,width=\linewidth,height=0.9\linewidth}}&
\mbox{\epsfig{file=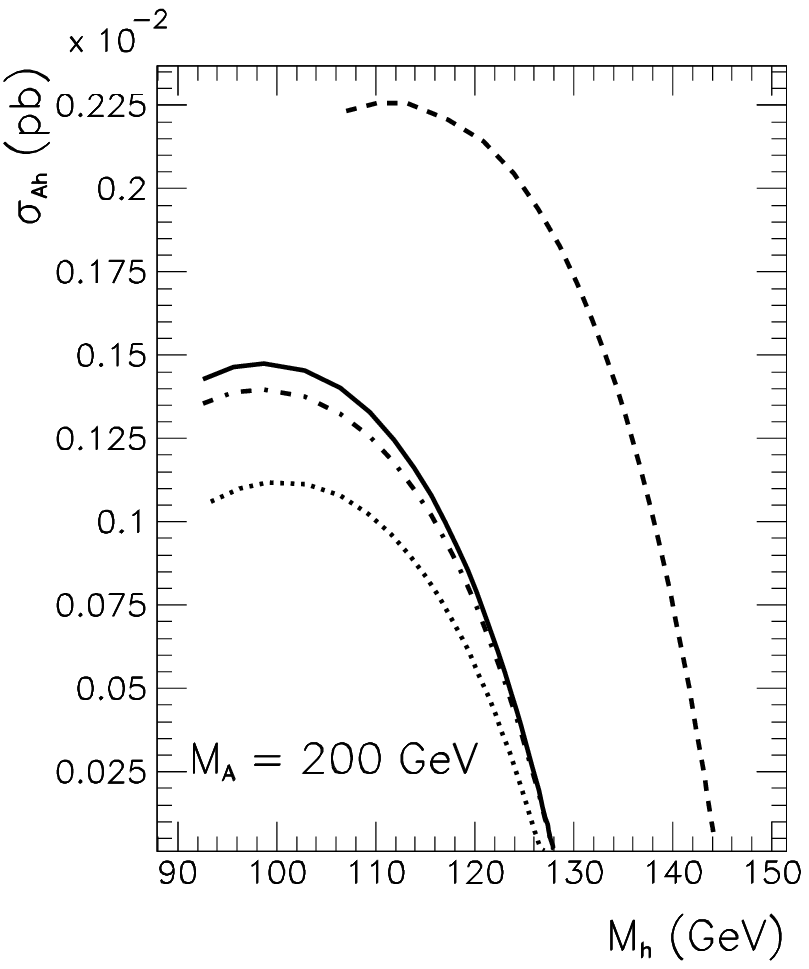,width=\linewidth,height=0.9\linewidth}}\\
\end{tabular}
\vskip -5mm 
\caption{$\sigma_{Zh}$ and $\sigma_{Ah}$ as a function of
  $\Mh$ at $\sqrt{s} = 500 \gev$, for two values of $\MA$, $\Xt/\msusy
  = 2$, and the parameters of Table~\ref{tab:par}. The solid (dot-dashed)
  line represents the \twol\ FD result including (excluding) box
  contributions, the dotted line shows the RG $\aeff$ approximation and the
  dashed line shows the \onel\ FD result.}
\label{fig:sigma_mh_s500}
\end{center}
\end{figure}
%%%%%%%%%%%% F I G U R E %%%%%%%%%%%%%%%%%%%%%%%%%%%%%%%%%%%%%

In \reffi{fig:sigma_mh_s500} the cross sections for the
Higgs-strahlung process and the associated production are shown for a
typical Linear Collider energy, $\sqrt{s}=500$ GeV~\cite{teslacdr}, in
the maximal-mixing scenario (in the no-mixing scenario similar results
have been obtained). In addition to the previous plots we also show
the result for the \twol\ FD calculation where the box contributions
have not been included, in order to point out their relative
importance for high-energy collisions.  For $\sqrt{s} = 500 \gev$ the
differences between the FD result and the RG $\aeff$ approximation
are larger than in the low-energy
scenario, an effect that is more pronounced for the higher value of
$\MA$.  For $\MA = 200 \gev$, typically they are of the order of
10-15\% for $\sigma_{Zh}$ and even up to 25\% for $\sigma_{Ah}$ (for
$\Mh \gsim 90 \gev$).
The difference between the \twol\ and \onel\ FD result can be
sizable. The \twol\ result for $\si_{Zh}$ is in general larger than
the \onel\ value, again increasing with $\MA$, where for $\MA =
200 \gev$ the difference can amount up to 15\%. $\si_{Ah}$, on the 
other hand, is
decreased at the \twol\ level for $\MA = 200 \gev$ and the difference
may be sizable.

The box contributions become more important for higher $\sqrt{s}$ and
change the total cross section by 5-10\%.  This result remains
unchanged even if sleptons are significantly heavier than
$M_{\tilde{l}}=300$ GeV used in our numerical analysis, as the
dominant contributions to box diagrams are given by $W$ and Higgs
boson exchanges~\cite{boxes,higgsprod}, which do not depend on
$M_{\tilde{l}}$.  Also, one should recall that box contributions
lead to an angular distribution of the final-state particles
different from the effective Born approximation and thus give much larger
corrections to the differential rather than to the total cross
section, at least for some range of the scattering angle.  Therefore,
the box diagrams have a significant effect at Linear Collider energies
and thus have to be included.  The same conclusions can be drawn for
$\sqrt{s} = 500 \gev$ in the no-mixing scenario, which we do not show
here. The differences between the FD result and the RG $\aeff$
approximation are only slightly smaller than in the maximal-mixing case.

%%%%%%%%%%%% F I G U R E %%%%%%%%%%%%%%%%%%%%%%%%%%%%%%%%%%%%%
\begin{figure}[htb!]
\begin{center}
\begin{tabular}{p{0.48\linewidth}p{0.48\linewidth}}
\mbox{\epsfig{file=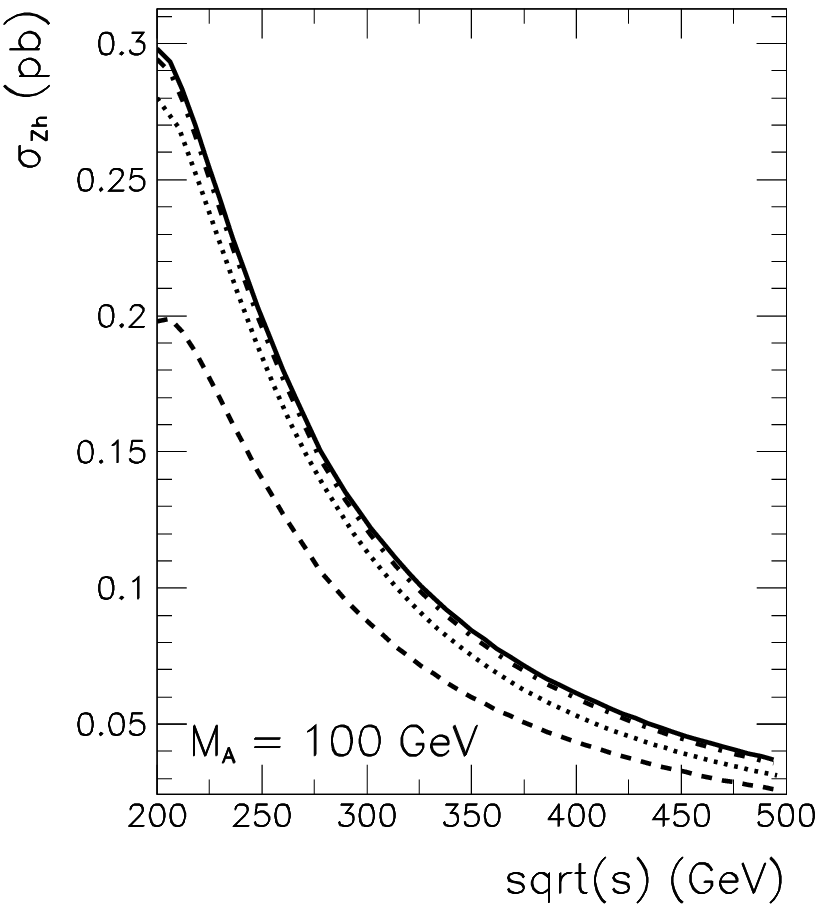,width=\linewidth,height=0.95\linewidth}}&
\mbox{\epsfig{file=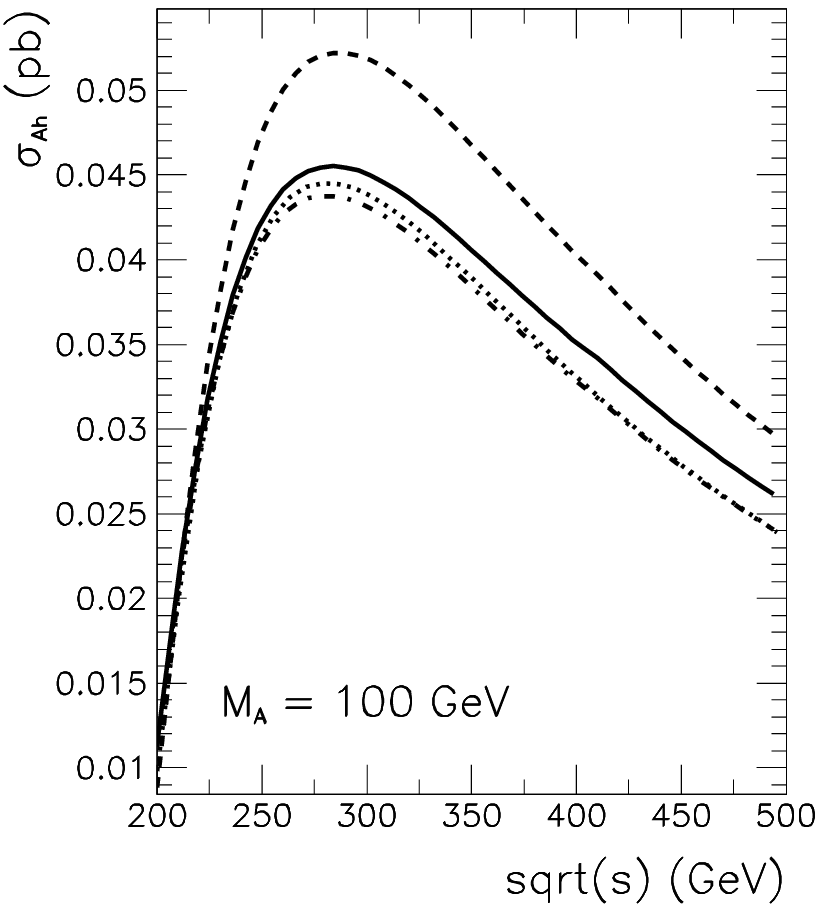,width=\linewidth,height=0.95\linewidth}}\\
\mbox{\epsfig{file=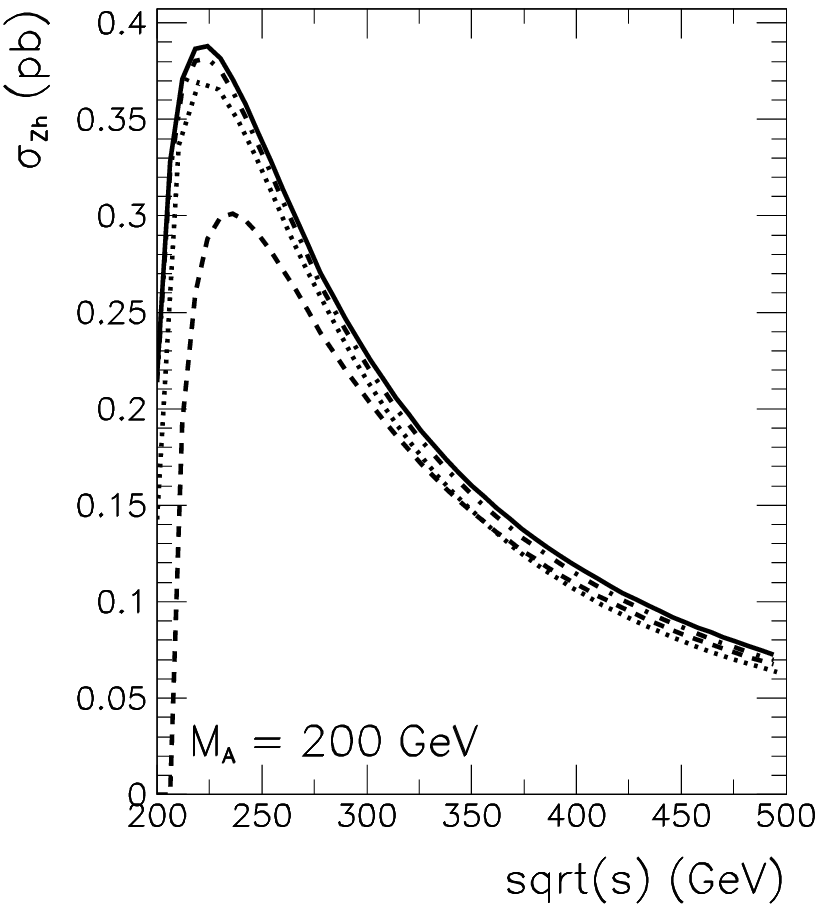,width=\linewidth,height=0.95\linewidth}}&
\mbox{\epsfig{file=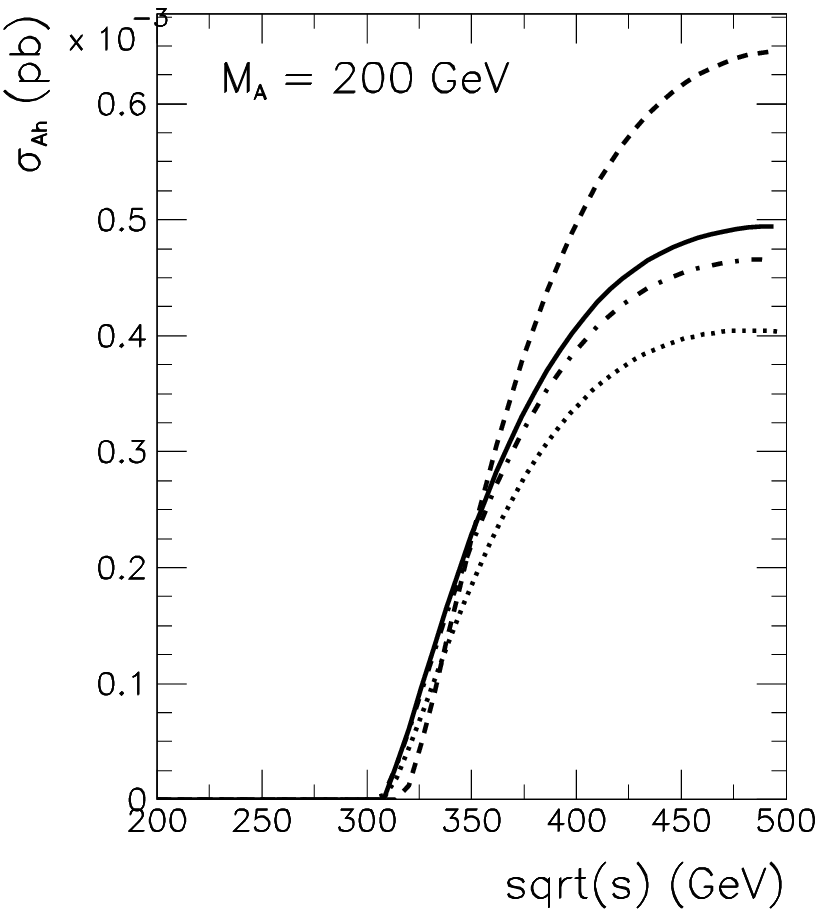,width=\linewidth,height=0.95\linewidth}}\\
\end{tabular}
\vskip -5mm 
\caption{$\sigma_{Zh}$ and $\sigma_{Ah}$ as a function of
  $\sqrt{s}$ for $\tb = 5$, $\Xt = 0$, and the parameters of
  Table~\ref{tab:par}.  The solid (dot-dashed) line represents the
  \twol\ FD result including (excluding) box contributions, the dotted
  line shows the RG $\aeff$ approximation and the dashed line shows the
  \onel\ FD result.}
\label{fig:sigma_s_tb5}
\end{center}
\end{figure}
%%%%%%%%%%%% F I G U R E %%%%%%%%%%%%%%%%%%%%%%%%%%%%%%%%%%%%%

%%%%%%%%%%%% F I G U R E %%%%%%%%%%%%%%%%%%%%%%%%%%%%%%%%%%%%%
\begin{figure}[htb!]
\begin{center}
\begin{tabular}{p{0.48\linewidth}p{0.48\linewidth}}
\mbox{\epsfig{file=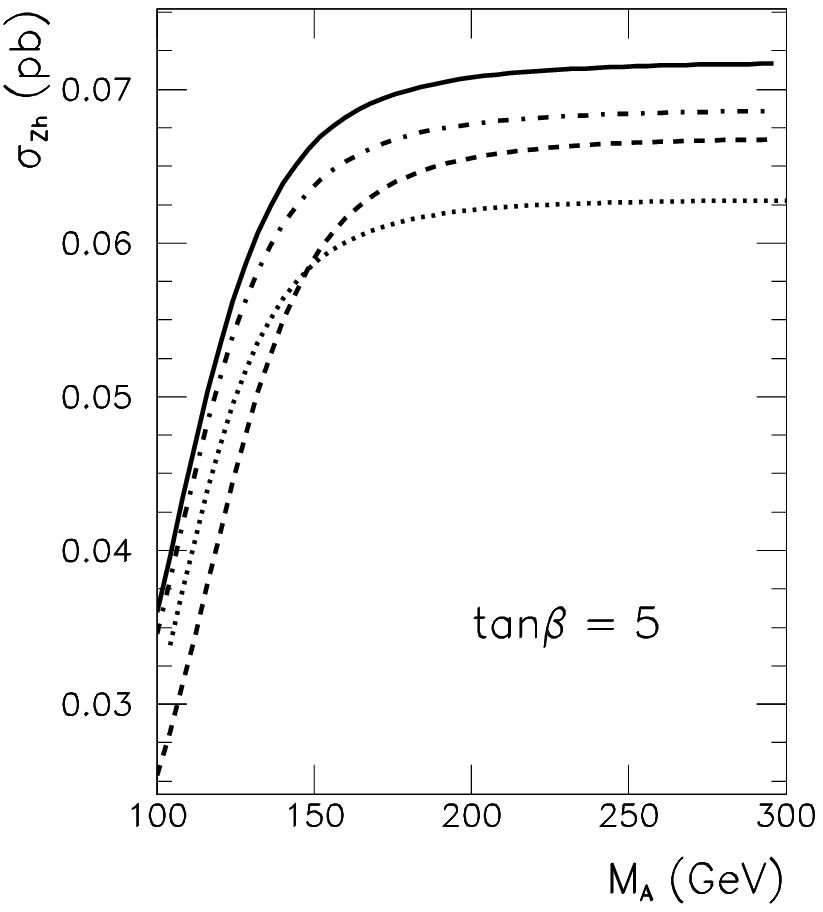,width=\linewidth,height=0.9\linewidth}}&
\mbox{\epsfig{file=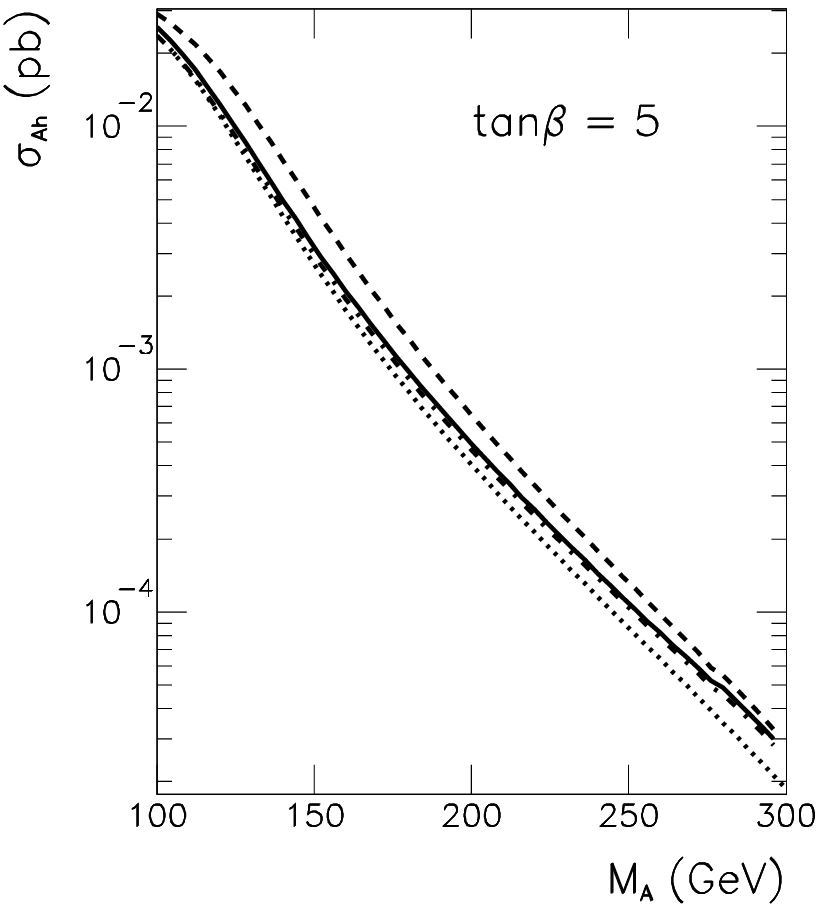,width=\linewidth,height=0.9\linewidth}}\\
\mbox{\epsfig{file=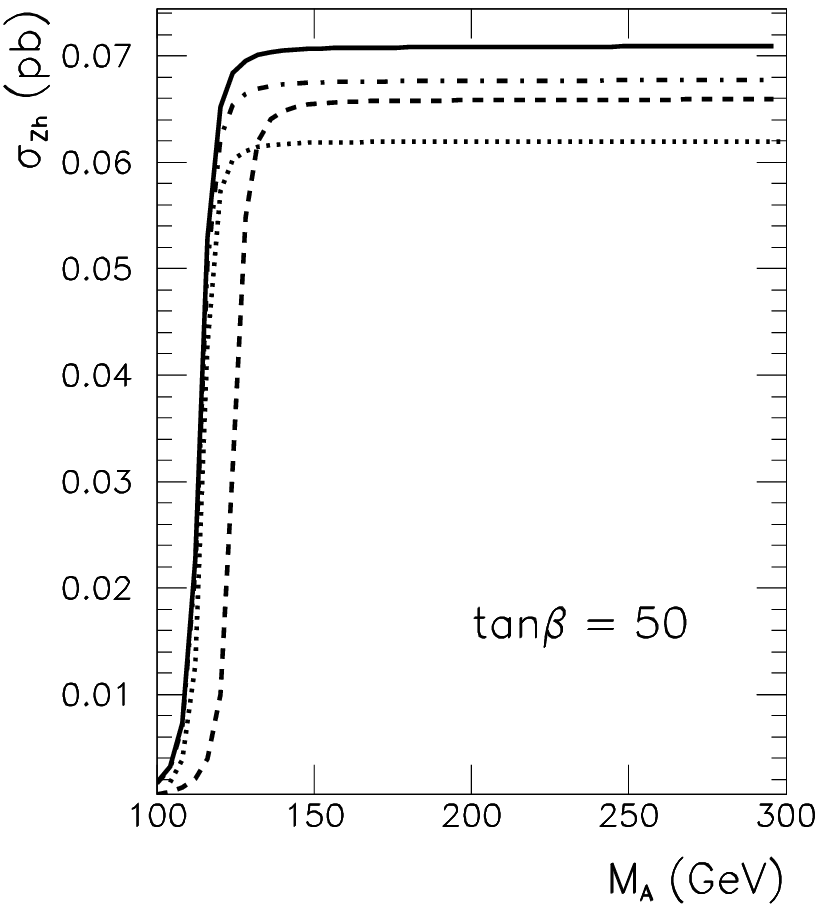,width=\linewidth,height=0.9\linewidth}}&
\mbox{\epsfig{file=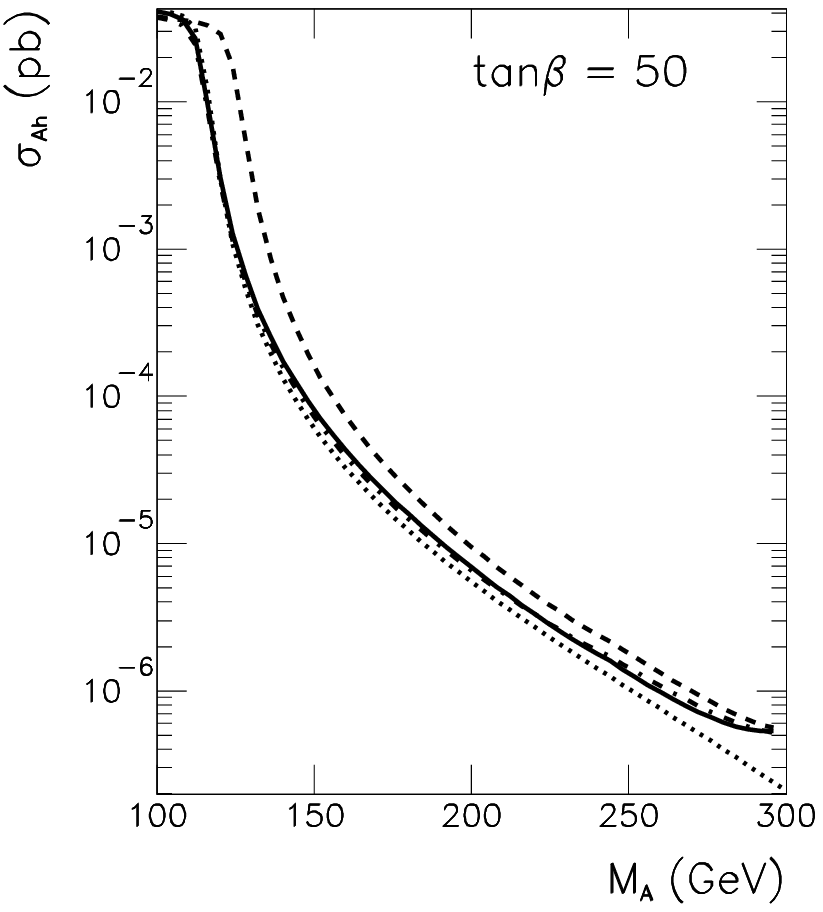,width=\linewidth,height=0.9\linewidth}}\\
\end{tabular}
\vskip -5mm 
\caption{$\sigma_{Zh}$ and $\sigma_{Ah}$ as a function of
  $\MA$ at $\sqrt{s} = 500 \gev$, shown for $\tb = 5$ and $\tb = 50$,
  $\Xt = 0$, and the parameters of Table~\ref{tab:par}.  The solid
  (dot-dashed) line represents the \twol\ FD result including
  (excluding) box contributions, the dotted line shows the RG $\aeff$
  approximation and the dashed line shows the \onel\ FD result.}
\label{fig:sigma_MA}
\end{center}
\end{figure}
%%%%%%%%%%%% F I G U R E %%%%%%%%%%%%%%%%%%%%%%%%%%%%%%%%%%%%%

In \reffi{fig:sigma_s_tb5} the results for $\sigma_{Zh}$ and
$\sigma_{Ah}$ are shown as a function of $\sqrt{s}$ in the no-mixing
scenario for $\MA = 100, 200 \gev$.  Besides the obvious kinematical
drop-off of the cross sections, one can observe that the relative
differences between the FD two-loop result and the RG $\aeff$
approximation  grow with~$\sqrt{s}$. The differences
remain almost constant or even increase slowly in absolute terms,
whereas the full cross sections decrease.  $\sigma_{Ah}$ becomes very
small for large $\MA$, as can be seen in more detail in
\reffi{fig:sigma_MA}. There we show the dependence of
$\sigma_{Zh}$ and $\sigma_{Ah}$ on $\MA$ in the no-mixing scenario for
$\tb = 5$ and $\tb=50$.  For $\sigma_{Zh}$, the $A$ boson decouples
quickly; the dependence on $M_A$ becomes very weak for
$\MA \gsim 250 \gev$, when $\sigma_{Zh}$ is already practically
constant (compare e.g.~\citere{mhiggslong}).  In the same limit,
$\sigma_{Ah}$ goes quickly to zero due to suppression of the effective
$ZhA$ coupling, which is $\sim\cos(\aeff - \be)$; 
also the kinematical suppression
plays a role, but this becomes significant only for sufficiently large
$\MA$, $\MA>350 \gev$.  For large $\tb$ the decoupling of $M_A$ is
even more rapid. The differences between the FD two-loop result and
the RG $\aeff$ approximation
for the Higgs-strahlung cross section tend also to a constant, but
they increase with $\MA$ for the associated production. The latter can
be explained by the growing relative importance of 3- and 4-point
vertex function contributions compared to the strongly suppressed
Born-like diagrams. As can be seen from~\reffi{fig:sigma_MA}, for
$\tb=50$ and $\MA\geq 300 \gev$ the FD two-loop result is almost an order of
magnitude larger than the result of the RG $\aeff$ approximation, and 
starts to saturate. This can be
attributed to the fact that the (non-decoupling) vertex and box
contributions begin to dominate the cross section value. However, such
a situation occurs only for very small $\sigma_{Ah}$ values,
$\sigma_{Ah} \approx 10^{-3}$~fb, below the expected experimental LC
sensitivities.

%\bigskip
%In order to explore the full effect of our calculation on ongoing
%Higgs analyses at LEP2 or searches at future colliders, a comparison
%with currently available codes, e.g.\ HZHA~\cite{hzha}, has to be
%performed. 
%Therefore our results are currently implemented in
%HZHA~\cite{feynhiggsXSinHZHA}. The effects, including 
%initial-state QED corrections as well as
%finite-width effects (allowing for off-shell decays of the Higgs and
%the $Z$~boson), will be discussed in a forthcoming
%publication~\cite{eehZhA2}. 

%%%%%%%%%%%%%%%%%%%%%%%%%%%%%%%%%%%%%%%%%%%%%%%%%%%%%%%%%%%%%%
%%%%%%%%%%%%%%%%%%%%%%%%%%%%%%%%%%%%%%%%%%%%%%%%%%%%%%%%%%%%%%

%%%%%%%%%%%%%%%%%%%%%%%%%%%%%%%%%%%%%%%%%%%%%%%%%%%%%%%%%%%%%%
%%%%%%%%%%%%%%%%%%%%%%%%%%%%%%%%%%%%%%%%%%%%%%%%%%%%%%%%%%%%%%

\section{Conclusions}
\label{sec:concl}

Using the Feynman-diagrammatic approach we have calculated the
production cross sections for \eetohZhA. The Higgs-boson propagator
corrections have been evaluated, including besides the full \onel\
result also the dominant and subdominant \twol\ corrections. In
addition we have included also the full set of \onel\ vertex and box
corrections.

We have also investigated an improved Born approximation based on the 
effective mixing angle, $\aeff$, in the neutral $\cp$-even Higgs sector.
We have shown analytically that this approximation corresponds to taking
into account the Higgs-boson propagator corrections with neglected
external momentum at one-loop and two-loop order. 

While the previous results available in the literature neglected the
large two-loop corrections in the case of the Feynman-diagrammatic
approach and were restricted to an improved Born approximation in the
case of the renormalization-group improved one-loop Effective Potential
approach, our new result combines the dominant two-loop corrections with the
complete Feynman-diagrammatic one-loop result and thus represents the
presently most precise prediction for the production cross sections for
\eetohZhA.

Specifically we have numerically analyzed the effect of the two-loop
contributions incorporated in our result, which turned out to be
sizable. We have furthermore compared our full 
Feynman-diagrammatic results for 
$\sigma_{Zh}$ and $\sigma_{Ah}$ with the approximation where $\aeff$ and 
$\Mh$ are evaluated within the renormalization-group improved one-loop
Effective Potential approach (RG $\aeff$ approximation).
For LEP2 energies, $\sqrt{s} = 206 \gev$, the difference between the
Feynman-diagrammatic result and the RG $\aeff$ approximation stays
mostly at the 
per cent level in the parameter space allowed by the LEP2 exclusion
limits for $\Mh$.
At energies reachable at an $e^+e^-$ linear collider, e.g.\
$\sqrt{s} = 500 \gev$, the difference between the Feynman-diagrammatic
result and the RG $\aeff$ approximation can reach
10-15\% for $\sigma_{Zh}$ or even 25\% for $\sigma_{Ah}$.
The box contributions play an important role for high $\sqrt{s}$ and
can amount up to 10\%.
Therefore in a precision analysis for a high
energy \epem\ collider the \twol\ propagator corrections as well as 
the complete diagrammatic one-loop contributions should be included%
\footnote{
A public Fortran code for the production cross sections is in
preparation~\cite{feynhiggs}.
}%
.

%%%%%%%%%%%%%%%%%%%%%%%%%%%%%%%%%%%%%%%%%%%%%%%%%%%%%%%%%%%%%%
%%%%%%%%%%%%%%%%%%%%%%%%%%%%%%%%%%%%%%%%%%%%%%%%%%%%%%%%%%%%%%

\section*{Acknowledgements}

Parts of the calculation have been performed on the QCM cluster at the
University of Karlsruhe, supported by the Deutsche
Forschungsgemeinschaft (Forschergruppe ``Quantenfeldtheorie und
Computeralgebra''). 
This work has been supported in part by the Foundation for
Polish-German Collaboration grant number 3310/97/LN and by the Polish
Committee of Scientific Research under the grant number 2~P03B~052~16,
1999-2000 (S.H. and J.R.).  
We express our gratitude to K.~Desch, P.~Janot and A.~Quadt for 
helpful discussions.

%%%%%%%%%%%%%%%%%%%%%%%%%%%%%%%%%%%%%%%%%%%%%%%%%%%%%%%%%%%%%%
%%%%%%%%%%%%%%%%%%%%%%%%%%%%%%%%%%%%%%%%%%%%%%%%%%%%%%%%%%%%%%

\newpage
\appendix

\renewcommand{\thesection}{\setcounter{equation}{0}Appendix~}
\renewcommand{\theequation}{\Alph{section}.\arabic{equation}}

\section{Explicit expressions for the cross sections}
\label{sec:app}

Performing the sum over polarization states in eq.~(\ref{eq:cr_sum_mat}) one
gets for the associated scalar and pseudoscalar production:
\BEA
{\cal A}_{1PS}^{i}=-~{\cal A}_{2PS}^{i}={e^2 \over 8} {\lambda}^2 \KL
s,M^2_A,M^2_{\Hi} \KR \KL
\left|a^{i}_{PS}\right|^2+\left|b^{i}_{PS}\right|^2 \KR ,
\EEA
where
\BEA
a^{i}_{PS} &=& \hat c_V \KL 2\tilde V^{(0)i1}_{ZPS}+\tilde
F_P^{i1}+\tilde F_S^{i1} \KR -2 \tilde V^{(0)i1}_{ZPS}{\hat\Sigma^T_{\ga
Z}(s)\over D_{\ga}(s)} + \KL {\tilde G}_P^{i1} + {\tilde G}_S^{i1} \KR
{D_Z(s)\over D_{\ga}(s)} , \nonumber\\
b^{i}_{PS} &=& \hat c_A \KL 2\tilde V^{(0)i1}_{ZPS}+\tilde
F_P^{i1}+\tilde F_S^{i1} \KR ,
\EEA
and $\la$ as given in \refeq{def:lambda}.
The corresponding expressions for the Higgs-strahlung process are 
more complicated:
\BEA
{\cal A}_{2ZS}^i &=& -{e^2{\lambda}^2 \KL s,M^2_Z,M^2_{\Hi} \KR \over
32M_Z^2} \times \nonumber\\ 
&\times& \KKL \left| 2 a^i_{ZS} - \KL s+M^2_Z-M^2_{\Hi} \KR b^i_{ZS}
\right|^2 - 4 s M_Z^2 \left| b^i_{ZS}\right|^2 + \right. \nonumber \\
&+&\left. \left| 2 c^i_{ZS}-
\KL s+M^2_Z-M^2_{\Hi} \KR d^i_{ZS}\right|^2 - 4 s M_Z^2
\left| d^i_{ZS} \right|^2 \KKR ,
\EEA
\BEA
{\cal A}_{1ZS}^i = e^2 s \KL
\left|a^i_{ZS}\right|^2+\left|c^i_{ZS}\right|^2 \KR - {\cal A}_{2ZS}^i ,
\EEA
where
\BEA
a^i_{ZS} &=& \hat c_V \KL \tilde V^{(0)i}_{ZZS}+\tilde F_1^i \KR -
\tilde V^{(0)i}_{ZZS} {\hat\Sigma^T_{\ga Z}(s)\over D_{\ga}(s)} +
{\tilde G}_1^i {D_Z(s)\over D_{\ga}(s)} , \nonumber\\
b^i_{ZS} &=& \hat c_V  \KL \tilde F_3^i -
\tilde F_4^i \KR  +  \KL {\tilde G}_3^i - {\tilde G}_4^i \KR  
{D_Z(s)\over D_{\ga}(s)} , \nonumber\\
c^i_{ZS} &=& \hat c_A \KL \tilde V^{(0)i}_{ZZS}+\tilde F_1^i \KR ,
\nonumber\\
d^i_{ZS} &=& \hat c_A \KL \tilde F_3^i - \tilde F_4^i \KR .
\EEA 

The quantities $\tilde F$ and $\tilde G$ denote effective 
Higgs--gauge-boson vertex form factors.  They can be defined as follows.  At the
tree level the relevant Higgs-boson vertices read (the assignment of
momenta is given in \refse{subsec:crosssection}):
\BEA
  V^{(0)\mu\nu i}_{ZZS}=iV^{(0)i}_{ZZS}~g^{\mu\nu} ,
\EEA
\BEA
  V^{(0)\mu ij}_{ZPS}=V^{(0)ij}_{ZPS} \KL p~-~q \KR ^{\mu} ,
\EEA
where $V^{(0)i}_{ZZS}$ and $V^{(0)ij}_{ZPS}$ can be written in
matrix form as (index $j$ numerates CP-odd Higgs bosons: $A\equiv
P_1$, $G\equiv P_2$):
\BEA
V^{(0)i}_{ZZS}={eM_Z\over s_W c_W} \KL 
\begin{array}{r}
\cos(\al-\be)\\
-\sin(\al-\be)
\end{array}
 \KR ,
\EEA
\BEA
V^{(0)ij}_{ZPS}=- {e\over 2s_Wc_W} \KL 
\begin{array}{rr}
\sin(\al-\be)&\cos(\al-\be)\\
\cos(\al-\be)&-\sin(\al-\be)
\end{array}
 \KR .
\EEA

After including \onel\ vertex corrections the renormalized vertices
have the form:
\BEA
V^{\mu ij}_{ZPS}= p^{\mu} \KL V^{(0)ij}_{ZPS}+\hat F_P^{ij} \KR 
-q^{\mu} \KL V^{(0)ij}_{ZPS}+\hat F_S^{ij} \KR ,
\label{eqn:zpsfull}
\EEA
\BEA
V^{\mu\nu i}_{ZZS}=i \KKL g^{\mu\nu} \KL V^{(0)i}_{ZZS}+\hat F^i_1 \KR
+p^{\mu}p^{\nu} \hat F^i_2+q^{\mu}q^{\nu} \hat F^i_3+
p^{\mu}q^{\nu}\hat F^i_4+q^{\mu}p^{\nu}\hat F^i_5 \KKR ,
\label{eqn:zzsfull}
\EEA
where $\hat F_a$'s are the renormalized vertex form factors.

Next, we define the respective quantities with tilde which are
obtained by inclusion of the \onel\ corrections on the external
lines. For instance, the complete vertices read (in the following: $i'
= 3 - i , j' = 3 -j$.):
\BEA
\tilde V^{ij}_{ZPS} &=&  \KL \Zext_{\rS i} 
\Zext_{\rP j} \KR ^{1/2} \KL  V^{ij}_{ZPS} + 
\Zmix_{\rS i} V^{i'j}_{ZPS} + \Zmix_{\rP j}
V^{ij'}_{ZPS} + \Zmix_{\rS i} \Zmix_{\rP j}
V^{i'j'}_{ZPS}  \KR ,
\label{eqn:vpstilde}
\EEA
and
\BEA
%\tilde V^{i}_{ZZS} =  \KL {\cal Z}_Z^{\rm ext} \Zext_{Si} \KR ^{1/2}
% \KL V^{i}_{ZZS} + \Zext_{S_{mix}i}V^{i'}_{ZZS} \KR 
\tilde V^{i}_{ZZS} =  \KL \Zext_Z \Zext_{\rS i} \KR ^{1/2}
 \KL V^{i}_{ZZS} + \Zmix_{\rS i}V^{i'}_{ZZS} \KR ,
\label{eqn:vzstilde}
\EEA
where
\BEA
\Zext_Z= \re \KL 1+{\partial\over\partial q^2}\left.
\hat\Sigma^T_{Z}(q^2)\right|_{q^2 = \MZ^2} \KR^{-1} .
\EEA
The vertices $\tilde V_{ZPS}, \tilde V_{ZZS}$ can be decomposed into
$\tilde V^{(0)}_{ZPS}, \tilde F_P, \tilde F_S$ etc. (as in
eq.~(\ref{eqn:zpsfull},\ref{eqn:zzsfull})) where e.g. $\tilde F^i_1 =
\KL {\cal Z}_Z^{\rm ext} \Zext_{\rS i} \KR ^{1/2} \KL F^i_1 
+ \Zmix_{\rS i}F^{i'}_1 \KR $ etc.

To include photon exchange in the $s$-channel one needs to know also
the renormalized vertices $V_{\ga PS}$ and $V_{\ga ZS}$ (which vanish at
the tree level):
\BEA
V^{\mu ij}_{\ga PS}=p^{\mu}\hat G_P^{ij}-q^{\mu}\hat G_S^{ij} ,
\EEA
\BEA
V^{\mu\nu i}_{\ga ZS}=i \KKL g^{\mu\nu}\hat G^i_1 +p^{\mu}p^{\nu}
\hat G^i_2+q^{\mu}q^{\nu} \hat G^i_3+ p^{\mu}q^{\nu} \hat
G^i_4+q^{\mu}p^{\nu} \hat G^i_5 \KKR .
\EEA
The vertices $\tilde V_{\ga PS}, \tilde V_{\ga ZS}$ are defined
similarly as shown in eqs.~(\ref{eqn:vpstilde},~\ref{eqn:vzstilde}).

The explicit expressions for the \onel\ corrections to the vertices
$V_{ZPS}$, $V_{ZZS}$, $V_{\ga PS}$ and $V_{\ga ZS}$ can be found in
Ref.~\cite{CPR}.

%%%%%%%%%%%%%%%%%%%%%%%%%%%%%%%%%%%%%%%%%%%%%%%%%%%%%%%%%%%%%%
%%%%%%%%%%%%%%%%%%%%%%%%%%%%%%%%%%%%%%%%%%%%%%%%%%%%%%%%%%%%%%

\end{document}